\documentclass[namedreferences]{solarphysics}
\usepackage[hyperref,optionalrh,solaromanenum]{spr-sola-addons} % For Solar Physics
\usepackage{graphicx}                    % For eps figures, newer & more powerfull
\usepackage{color}                       % For color text: \color command
\usepackage{breakurl}                         % For breaking URLs easily trough lines
\usepackage{lscape}
\usepackage{overpic}
\begin{document}
%\linenumbers
\begin{article}

\begin{opening}

%\title{Article title}
\title{Development of a Method for Determining the Search Window for Solar Flare Neutrinos}

%%%%%%%%%%%%%%%%%%%%%%%%%%%%%%%%%%%%%%%%%%%%%%%%%%%
%% Authors Names
%
% \author[addressref={},corref,email={}]{\inits{}\fnm{}\lnm{}\orcid{}}
\author[addressref={kamioka},corref,email={okamoto@km.icrr.u-tokyo.ac.jp}]{\inits{}\fnm{}\lnm{K.~Okamoto}\orcid{0000-0002-4280-6593}}

\author[addressref={kobe},corref,email={ynakano@phys.sci.kobe-u.ac.jp}]{\inits{}\fnm{}\lnm{Y.~Nakano}\orcid{0000-0003-1572-3888}}

\author[addressref={nagoya}]{\inits{}\fnm{}\lnm{S.~Masuda}}

\author[addressref={nagoya,KM}]{\inits{}\fnm{}\lnm{Y.~Itow}\orcid{0000-0002-8198-1968}}

\author[addressref={nagoya2}]{\inits{}\fnm{}\lnm{M.~Miyake}}

\author[addressref={kashiwa}]{\inits{}\fnm{}\lnm{T.~Terasawa}}

\author[addressref={okayama}]{\inits{}\fnm{}\lnm{S.~Ito}}

\author[addressref={kamioka,ipmu}]{\inits{}\fnm{}\lnm{M.~Nakahata}}

%%%%%%%%%%%%%%%%%%%%%%%%%%%%%%%%%%%%%%%%%%%%%%%%%%%
%% Runningheads
%
\runningauthor{K.~Okamoto~\textit{et al}.}
\runningtitle{Determining the Search Window for Solar Flare Neutrinos}

%%%%%%%%%%%%%%%%%%%%%%%%%%%%%%%%%%%%%%%%%%%%%%%%%%%
%% Affilations
%% id shold be the same with \author addressref value.
\address[id={kamioka}]{Kamioka Observatory, Institute for Cosmic Ray Research, The University of Tokyo, Kamioka, Gifu 506-1205, Japan}
\address[id={kobe}]{Department of Physics, Kobe University, Kobe, Hyogo 657-8501, Japan}
\address[id={nagoya}]{Institute for Space-Earth Environmental Research, Nagoya University, Nagoya, Aichi 464-8601, Japan}
\address[id={KM}]{Kobayashi-Maskawa Institute for the Origin of Particles and the Universe, Nagoya University, Nagoya, Aichi 464-8602, Japan}
\address[id={nagoya2}]{Department of Physics, Nagoya University, Nagoya, Aichi 464-8602, Japan}
\address[id={kashiwa}]{Institute for Cosmic Ray Research, The University of Tokyo, Kashiwa, Chiba 277-8582, Japan}
\address[id={okayama}]{Faculty of Science, Okayama University, Okayama, Okayama 700-8530, Japan}
\address[id={ipmu}]{Kavli Institute for the Physics and Mathematics of the Universe (WPI), The University of Tokyo Institutes for Advanced Study, The University of Tokyo, Kashiwa, Chiba 277-8583, Japan}
%%%%%%%%%%%%%%%%%%%%%%%%%%%%%%%%%%%%%%%%%%%%%%%%%%%
%%% Abstract
\begin{abstract}

Neutrinos generated during solar flares remain elusive. However, after 50~years of discussion and search, the potential knowledge unleashed by their discovery keeps the search crucial. Neutrinos associated with solar flares provide information on otherwise poorly known particle acceleration mechanisms during solar flare. For neutrino detectors, the separation between atmospheric neutrinos and solar flare neutrinos is technically encumbered by an energy band overlap. To improve differentiation from background neutrinos, we developed a method to determine the temporal search window for neutrino production during solar flares. Our method is based on data recorded by solar satellites, such as {\it Geostationary Operational Environmental Satellite} (GOES), {\it Reuven Ramaty High Energy Solar Spectroscopic Imager} (RHESSI), and GEOTAIL. In this study, we selected 23 solar flares above the X5.0 class that occurred between 1996 and 2018. We analyzed the light curves of soft X-rays, hard X-rays, $\gamma$-rays, line $\gamma$-rays from neutron capture as well as the derivative of soft X-rays. The average search windows are determined as follows: $4,178$~s for soft X-ray, $700$~s for derivative of soft X-ray, $944$~s for hard X-ray~($100$--$800$~keV), $1,586$~s for line $\gamma$-ray from neutron captures, and $776$~s for hard X-ray~(above $50$~keV). This method allows neutrino detectors to improve their sensitivity to solar flare neutrinos.
\end{abstract}

%%%%%%%%%%%%%%%%%%%%%%%%%%%%%%%%%%%%%%%%%%%%%%%%%%%
%% Keywords
%
\keywords{Solar flare, Neutrino, $\gamma$-ray, X-ray, Neutron, Particle acceleration}

\end{opening}
%-------------------------------------------------

%%%%%%%%%%%%%%%%%%%%%%%%%%%%%%%%%%%%%%%%%%%%%%%%%%%
%% Sections
%
% \section{}%\label{s:?}

\section{Introduction} \label{sec:intro}

\subsection{Neutrino Production in Solar Flare}

A solar flare is an explosive event at the surface of the Sun and is characterized by a rapid increase in the radiative flux. Through a process of magnetic reconnection, a solar flare converts its magnetic energy into thermal and kinetic energies of charged particles~\citep{parker_connection}. A typical solar flare is estimated to release $10^{26}$--$10^{32}$~erg~\citep{ellison}. On the basis of optical observations, a typical energy release time scale is $100$--$1,000$~s depending on the wavelength range~\citep{kane}. The occurrence rate of flares is also well studied, as it is the frequency distribution as a function of energy released. This distribution can be described by an inverse power-law in the energy range above $10^{27}~$erg~\citep{occur1, occur2}.

Solar flares accelerate charged particles, such as electrons and protons, and carry non-thermal energy from the acceleration site to the energy loss site in the chromosphere. These accelerated particles reach relativistic velocities in a short time, becoming so-called energetic particles. Additional counterparts, such as high energy particles~\citep{explorer1, oso1, masuda_flare} and  electromagnetic waves, have been observed and recorded in several satellites, for example, the {\it Orbiting Solar Observatory} series~\citep{oso2}, the {\it Geostationary Operational Environmental Satellite}~(GOES)~series~\citep{goes_tech}, the {\it International Sun-Earth Explorer 3}~(ISEE 3)~\citep{isee3}, the {\it Solar Maximum Mission} (SMM)~\citep{smm}, the {\it Reuven Ramaty High Energy Solar Spectroscopic Imager} (RHESSI)~\citep{rhessi, rhessi_det}, the {\it Compton Gamma Ray Observatory}~\citep{cgro}, HINOTORI~\citep{hinotori_flare}, Yohkoh~\citep{yohkoh},  Konus-Wind~\citep{konus_det}, and the Fermi-LAT~({\it Large Area Telescope}) satellite~\citep{fermi_lat}.

Although several models have been proposed to explain the particle acceleration associated with solar flares~\citep{model1, model2, model3}, the acceleration mechanism remains unknown. As neutral particles, such as neutrons, neutrinos, X-ray photons, and $\gamma$-ray photons, are not affected by magnetic field between the Sun and the Earth, these particles have an important role to play in our understanding of both the location and the time profile of particle acceleration during solar flares.

In particular, neutrinos associated with solar flares~(hereafter solar flare neutrinos) attract significant attention in the field of astrophysics. If protons are accelerated during solar flares to sufficiently high energies~(more than $300$~MeV), pions are produced by collisions with other nuclei in the solar atmosphere~\citep{pi02,pi03}. Finally, neutrinos are generated via the decay of those charged pions such as atmospheric neutrinos. Because of this hadronic origin, solar flare neutrinos give information about the acceleration of protons as well as subsequent interactions in the solar chromosphere.

\subsection{Brief History of Neutrinos Associated with Solar Flare}

Because the process of neutrino production is the same as that of atmospheric neutrinos, the energy range of solar flare neutrinos overlaps with that of atmospheric neutrinos. Hence, a technical difficulty arises in separating atmospheric neutrinos from solar flare neutrinos in neutrino experiments such as Super-Kamiokande (Super-K)~\citep{sk_det}, IceCube~\citep{icecube}, and so on. For the last 60~years, neutrinos from solar flares have been experimentally sought by Homestake~\citep{homestake}, Kamiokande~\citep{kam_flare}, SNO~\citep{sno_flare} and Borexino experiments~\citep{borexino} but remain unidentified.

In 1988, the Homestake experiment reported an excess of neutrino events when energetic solar flares occurred. This observation suggested a relation between solar flares and the neutrino capture rate on $\mathrm{^{37}Cl}$~\citep{bahcall1, bahcall2} and proposed neutrino magnetic moment spin precession to explain the time variation of the capture rate~\citep{right1, right2}. However, the Kamiokande, SNO and Borexino experiments searched for neutrino signals using different solar flare samples and found no candidate signal related to solar flares.

Last several decades, some emission models of solar flare neutrinos have been theoretically discussed~\citep{wilson, boya1,boya2, farg1, farg2, kocha, take}. For example, \citealp{farg1} and \citealp{farg2} predicted a possibility of observing neutrinos from a large solar flare of energy greater than $10^{32}$~erg using Super-K and IceCube. On the other hand, \citealp{take} predicted no possibility of observing neutrinos in Hyper-Kamiokande~\citep{hk} by numerical simulations.

\subsection{Time Window for Neutrino Search}

As explained above, the separation of atmospheric neutrinos from solar flare neutrinos is technically difficult. Although atmospheric neutrinos are constantly generated, solar flare neutrinos are expected to be released only during the time scale of particle acceleration. Therefore, the signal-to-noise ratio for observing solar flare neutrinos can be improved by setting an appropriate search window.

The National Oceanic and Atmospheric Administration~(NOAA) maintains a list of solar flares observed by the GOES satellite. This list summarizes the date of each solar flare, its location, its class, its start and end times, and so on. This information can be used to estimate the time scale of solar flares. However it is questionable whether this time window is appropriate for solar flare neutrino searches, because the GOES satellite provides data relating only to the intensity of soft X-rays produced by thermal electrons. For this reason, the GOES data are not particularly useful in identifying the production time scale of solar flare neutrinos.

In the previous study, \citealp{window} proposed a method for determining the search window for solar flare neutrinos using $\gamma$-rays above $70$~MeV for neutral pion decays ($\pi^{0}\to2\gamma$) observed by the Fermi-LAT satellite~\cite{fermi_flare}. The appearance of neutral pions implies the generation of neutrinos in solar flares, because neutral pions in solar flares are produced simultaneously with the charged pions. Therefore, it is natural to set the search window for solar flare neutrinos based on occurrence time frame of $\gamma$-ray emission caused by neutral pions. However, as the Fermi-LAT satellite was launched only in 2008, the earlier period is not covered and the observation of such $\gamma$-ray emission is rare~\citep{kuz}. Notably, this restriction prevents the use of this method for the largest solar flare on record November 04, 2003, class X$28.0$~\citep{x28}.

For the above reasons, the appropriate time window for searching for solar flare neutrinos must be set using an alternative method. In this study, we analyzed not only soft X-rays but also $\gamma$-rays and hard X-rays because an accelerated proton during a solar flare also has a chance of generating $\gamma$-rays and neutrons. As mentioned above, $\gamma$-rays indicate the timing of proton interactions in the solar atmosphere, with neutrinos expected to be produced simultaneously. Therefore, timing information recorded by either $\gamma$-ray or X-ray instruments allows neutrino detectors to set the appropriate time window for solar flare neutrino searches.

This paper is organized as follows. In Section~\ref{sec:over}, we provide a brief overview of X-rays and $\gamma$-rays for solar flares. In Section~\ref{sec:method}, we describe a method for determining the time window to search for neutrinos from solar flares. In Section~\ref{sec:result}, we present the analysis result. In the final section, we discuss the conclusions and future prospects.

\section{Overview of Electromagnetic Observations Related to Accelerated Particles} \label{sec:over}

The measurement of electromagnetic wave energy allows the identification of its production and origin. The different mechanisms generate different kinds of brightening or dimming in the energy flux~\citep{kane}. Several observations of neutrons and electromagnetic waves associated with solar flares have been performed by space satellites since the 1960s. Alothough there are electromagnegic observations in Extreme Ultraviolet, radio and visible light range, we focus on X-ray and $\gamma$-ray observations to discuss neutrino emission in solar flares.

\subsection{Soft X-rays}

Soft X-rays are originally produced by bremsstrahlung from electrons in thermal motion. Because the energy of a particle accelerated by a solar flare is ultimately converted into thermal energy, the total intensity of the soft X-ray is used to represent the total energy released by the solar flare. A peak value for the soft X-ray flux in the range of $1$--$8~\mathrm{\AA}$~($1.5$--$12$~keV) observed by GOES is widely used as an indicator in the classification of solar flares. On the Basis of this classification, a solar flare whose peak value exceeds $1.0\times10^{-4}~\mathrm{W/m^{2}}$~(X1.0 class) is considered to be the largest order~\citep{goes_tech2}.

\subsection{Hard X-rays}

Hard X-rays are produced by bremsstrahlung from non-thermal electrons and the energy ranges from a few keV to MeV. Such non-thermal electrons are accelerated by the solar flare; hence they provide information about the energy distribution of the non-thermal electrons. Non-thermal electrons lose their energy by scattering in the solar atmosphere and eventually generate thermal bremsstrahlung. In contrast with soft X-rays, hard X-rays provide information about the electron acceleration site and its time frame \citep{Fletcher2011, Holman, kontar2011}.

\citet{neupert} found that the timing of the soft X-ray derivative maximum is almost the same as the peak time of the non-thermal microwave emissions in the impulsive phase of solar flares and interpreted that the heating may be the result of collisional losses by energetic accelerated electrons.
This is called the Neupert effect and can be applied in the case of non-thermal hard X-rays emissions instead of microwave emission~\citep{neupert_stat, goes_neupert}.

Using this effect, we can extract information about electron acceleration from soft X-ray data because a light curve of soft X-rays differentiated by time is expected to be similar to a light curve of hard X-rays. This method is useful to estimate a time profile of electron acceleration even when hard X-ray data are not available during a solar flare.

\subsection{$\gamma$-rays}

Because neutrinos are produced via hadronic decay, the timing of these interactions should be identified in order to determine the search window. The time profile of $\gamma$-rays is the most important in this study because an observation of $\gamma$-rays implies the acceleration and nuclear reactions of protons. In a solar flare, there are several different reactions produce $\gamma$-rays; $\gamma$-rays due to bremsstrahlung from high energy electrons, $\gamma$-rays due to electron-positron annihilation, line $\gamma$-rays due to neutron capture on hydrogen nuclei, and $\gamma$-rays due to de-excitation from nuclei, such as carbon~($\mathrm{^{12}C}$) and oxygen~($\mathrm{^{16}O}$)~\citep{carbon1, carbon}. Among these, line $\gamma$-rays due to neutron capture on hydrogen nuclei are the most important channel and are discussed below, along with $\gamma$-ray emission due to neutral pions.

\subsubsection{Line $\gamma$-ray from Neutron}

High energy protons accelerated by a solar flare collide with helium in the solar chromosphere and finally produce neutrons. Such neutrons undergo thermalization and ultimately produce $2.223$~MeV of $\gamma$-rays, so called line $\gamma$-rays, when they are captured on hydrogen nuclei. A line $\gamma$-ray is an indirect evidence of hadronic interactions including neutrons and high energy protons in a solar flare. Therefore, the determination of the search window should include analysis of the timing of line $\gamma$-rays, as their appearance indicates the acceleration of protons. However, in so doing we must take into account the delay time of about $100$~s from the generation of line $\gamma$-rays to the capture of produced neutrons on hydrogen~\citep{gan}.

As explained above, hard X-rays are produced by the acceleration of electrons, and line $\gamma$-rays are produced by hadronic interactions after the acceleration of protons~(i.e. ions). Shih {\it et al.,} 2009 found a linear correlation between the fluence of the $2.223$~MeV $\gamma$-ray and the fluence of the hard X-ray above $300$~keV in a large solar flare. Although it is unclear whether each acceleration happens at the same time and at the same location, this correlation implies a close relationship between the acceleration process of electrons and that of protons. Assuming this correlation, a timing delay due to neutron capture is expected in the line $\gamma$-ray flux in comparison with the hard X-ray flux.

\subsubsection{$\gamma$-rays from Pions}

Neutral pions~($\pi^{0}$) appear in proton-proton collisions when their energies exceed $300$~MeV. The observation of $\gamma$-ray between $70$~MeV and $100$~MeV produced via neutral pion decays~($\pi^{0}\to2\gamma$) implies the production of neutrinos during solar flares because neutral pions are produced simultaneously with the charged pions~\citep{kurt}. In fact, the Fermi-LAT satellite and Coronal-F satellite's SONG detector have both provided examples of $\gamma$-ray observations from neutral pions. In the case of the solar flare on October 28, 2003, SONG recorded a sharp $\gamma$-ray increase due to $\gamma$-ray emission from neutral pions. In this case, the emission lasted for at least $8$--$9$~min~\citep{kurt2, kuz}. On the contrary, Fermi-LAT reported that the $\pi^{0}$ emission after the solar flare on September 10, 2017 lasted for about 12 h~\citep{omodei}. On the basis of these observations, the expected solar flare neutrino rate at the Earth's surface may be estimated by numerical calculation. However, this estimation is beyond the scope of this paper, because the time profile of $\gamma$-ray emission remains unclear.

\section{Analysis Method} \label{sec:method}

To search for solar flare neutrinos, we used public data from solar satellites and computational resources of the Center for Integrated Data Science~(CIDAS) at Nagoya University. In this study, we analyzed the data recorded by three satellites; GOES, RHESSI and GEOTAIL. The main~(sub) targets for each satellite are summarized in Table~\ref{tbl:satellite}.

\begin{table}
\caption{Observation targets for each solar satellite. Circles~($\bigcirc$) show the main target whose energy and timing can be measured by the satellite's devices. Triangles~($\bigtriangleup$) indicate secondary target, hardly detectable by the satellite as direct measurements; the counting rate alone can indirectly imply target existence. Details are described in the later sections.} \label{tbl:satellite}
\begin{tabular}{ccccccc}
\hline
Satellite &Launch year & End year& Soft X-ray & Hard X-ray & Line $\gamma$-ray & $\gamma$-ray \\ \hline
GOES & 1975 & -- & $\bigcirc$ & $\bigtriangleup$ & -- & -- \\
RHESSI & 2002 & 2018 & $\bigcirc$ & $\bigcirc$ & $\bigcirc$ & $\bigcirc$ \\
GEOTAIL &1992 & -- & -- & $\bigtriangleup$ & $\bigtriangleup$ & $\bigtriangleup$ \\ \hline
\end{tabular}
\end{table}

Using CIDAS, we analyzed the GOES satellite's soft X-ray data, as well as the RHESSI satellite's hard X-ray and $\gamma$-ray data, as summarized in Table~\ref{tbl:energy_range}. For the analysis of GEOTAIL data, we checked the public data in CIDAS and then analyzed the corresponding raw data in order to recover hard X-rays above $50$~keV including soft $\gamma$-rays.

\begin{table}
\caption{Summary of satellites and energy ranges in this study. As described in the main text, GEOTAIL cannot measure the energy of neutral charged particles above $50$~keV. } \label{tbl:energy_range}
\begin{tabular}{ccc}
\hline
Type & Satellite & Energy range~(wave length) \\ \hline
Soft X-ray & GOES & $1.5$--$12$~keV~($1$--$8$~$\mathrm{\AA}$) \\
Hard X-ray & RHESSI & $100$--$800$~keV \\
Line $\gamma$-ray & RHESSI & $2.218$--$2.228$~MeV \\
Hard X-ray and soft $\gamma$-ray & GEOTAIL & Above $50$~keV \\ \hline
\end{tabular}
\end{table}

\subsection{Soft X-rays~(GOES)} \label{subsec:soft}

We used data obtained by the GOES satellite to determine the search window, because the GOES satellite series have continually monitored solar flares since 1975. Its long operation permits the use of the same indicator in comparisons between older and more recent flares.

For this study, we first registered all solar flares occurring from 1996 to 2018 in the NOAA list. Then, we selected the 23 solar flares whose peak soft X-ray intensity exceeds $5.0\times10^{-4}~\mathrm{W/m^{2}}$~(X5.0 class). The date, class, and location on the Sun~(active region, so called AR) for each selected flare are summarized in  Table~\ref{table:flares}.

\begin{table}
%  \begin{center}
    \caption{List of solar flares selected for this study. The data, class and active region~(AR) location are taken from a National Oceanic and Atmospheric Administration source~(NOAA, https:\slash{}\slash{}www.ngdc.noaa.gov\slash{}stp\slash{}space-weather\slash{}solar-data\slash{}solar-features\slash{}solar-flares\slash{}x-rays\slash{}goes\slash{}xrs\slash{}).} \label{table:flares}
%    \scalebox{0.7}{
    \begin{tabular}{ccc} \hline
      Date&Class&AR location \\ \hline
      1997 Nov. 6 & X9.5 & S18W63\\
      2000 Jul. 14& X5.7 & N22W07 \\
      2001 Apr. 2 & X20.0& S21E83 \\
      2001 Apr. 6 & X5.6 & S21E31 \\
      2001 Apr. 15& X14.4& S20W85 \\
      2001 Aug. 25& X5.8 & S17E34 \\
      2001 Dec. 13& X5.3 & N16E09  \\
      2002 Jul. 23& X5.1 & S13E72  \\
      2003 Oct. 23& X5.4 & S21E88 \\
      2003 Oct. 28& X17.2& S16E08  \\
      2003 Oct. 29& X10.0& S15W02 \\
      2003 Nov. 2 & X9.2 & S14W56  \\
      2003 Nov. 4 & X28.0& S19W83  \\
      2005 Jan. 20& X7.1 & N14W61  \\
      2005 Sep. 7 & X18.2& S11E77  \\
      2005 Sep. 8 & X5.4 & S12E75 \\
      2005 Sep. 9 & X6.2 & S12E67  \\
      2006 Dec. 5 & X9.0 & S07E68 \\
      2006 Dec. 6 & X6.5 & S05E64  \\
      2011 Aug. 9 & X7.4 & N14W69  \\
      2012 Mar. 7 & X5.4 & N22E12  \\
      2017 Sep. 6 & X9.4 & S09W34 \\
      2017 Sep. 10& X8.3 & S08W88 \\ \hline
    \end{tabular}
%	}%scale box
%  \end{center}
\end{table}

\begin{figure}
\centerline{\includegraphics[width=0.8\textwidth,clip=]{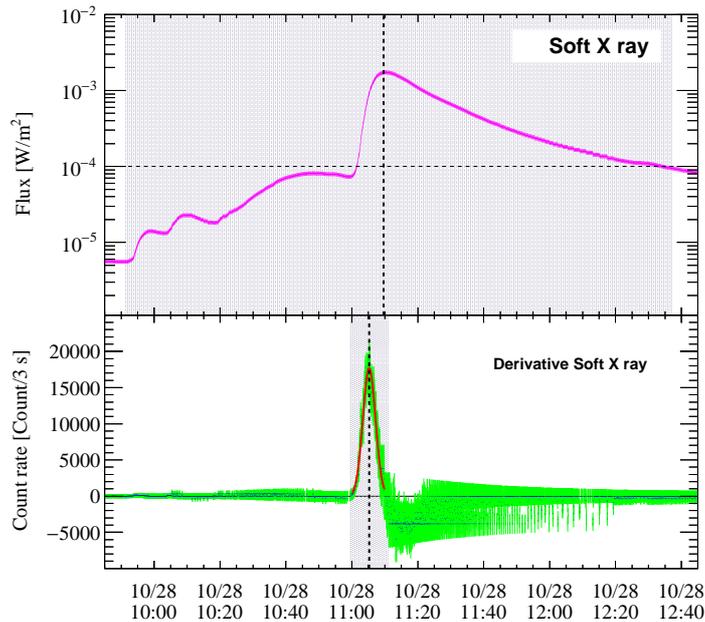}}
\caption{An example solar flare occurring on October 28, 2003, showing the soft X-ray light curve~(top) observed by the GOES satellite and the derivative of the flux of soft X-rays~(bottom). In the top panel, the horizontal axis is universal time and the vertical axis is the flux of soft X-rays. The black dotted horizontal~(vertical) line shows the flux of $1.0\times10^{-4}~\mathrm{W/m^{2}}$~(the peak timing of this channel). The region drawn in gray shows the time window selected in this study. In the bottom panel, the horizontal axis is universal time and the vertical axis is the derivative of the flux of soft X-rays with respect to time. The blue dots and the green region depict the data and their error, respectively. The red curve is the result of the Gaussian fitting and the vertical dotted line shows the peak timing of this channel. The gray region shows the time window selected by this study.}\label{fig:soft}
\end{figure}

A typical soft X-ray light curve observed by the GOES satellite is shown in the top panel of Figure~\ref{fig:soft}. In the NOAA list, the timing at which four consecutive flux increases occur was determined as the start time of each solar flare. The flare end time was typically determined as the timing at which the event's flux decline to half the value of its peak. However, this end time depends on the flare class which make unsuitable for our analysis. We therefore set the flare end time as the timing at which the flux declines to $1.0\times10^{-4}~\mathrm{W/m^{2}}$~(X1.0~class).

To extract information about light curve brightening correlated with particle acceleration, we took a calculation of the time derivative of the light curve as shown in the bottom panel of Figure~\ref{fig:soft}. In this analysis, we used the $1$--$8$~$\mathrm{\AA}$ band the same as used by \citealp{kurt}. For setting the time window, we first fitted the light curve with a Gauss function~($A\exp\{(t-t_{\mathrm{peak}})^{2}/\sigma^{2}\}$). We then selected the region of $[t_{\mathrm{peak}}-3\sigma,~t_{\mathrm{peak}}+3\sigma]$ as the search window.

\subsection{Hard X-ray~(RHESSI)} \label{subsec:hard}

The RHESSI satellite was launched on February 2002 to monitor solar flares. It records electromagnetic waves in the range of $3$~keV to $17$~MeV and measures continuous spectra from bremsstrahlung of high energy electrons, line $\gamma$-rays of 2.223 MeV, and $\gamma$-rays from $\mathrm{^{12}C}$ and $\mathrm{^{16}O}$ nuclei~\citep{rhessi}. Owing to its wide energy range, the RHESSI satellite provides important information for extracting acceleration time scales for electrons, protons, and ions.

In this study, we integrated the energy spectrum in the range of $100$--$800$~keV for hard X-ray analysis as shown in the top panel of Figure~\ref{fig:rhessi}a. We then determined the count rate within this energy range using the front detector on the RHESSI because the front detectors are sensitive to $100$--$300$~keV emission and they are dominant in the range of $100$--$800$~keV. To set the search window, we fitted the light curve via a combination of basic functions. For the fitting, we used constant value for the background before the brightening, a linear function for the impulsive phase of the light curve, and an exponential function for the gradual phase. The start time~($t_{\mathrm{start}}$) is defined as the time when the liner function intersects the background constant. The time window was determined by selecting the region where the exponential function is $2\sigma$ above the background constant. The standard deviation~($\sigma$) is obtained as standard deviation of constant background fitting. The end time~($t_{\mathrm{end}}$) is the last point satisfying this condition. In this way, the search window by hard X-ray is determined as [$t_{\mathrm{start}},~t_{\mathrm{end}}$].

\subsection{Line $\gamma$-rays~(RHESSI)}

\begin{figure}
\centerline{\hspace*{0.015\textwidth}
\begin{minipage}{0.49\hsize}
\centerline{\includegraphics[width=1.0\textwidth,clip=]{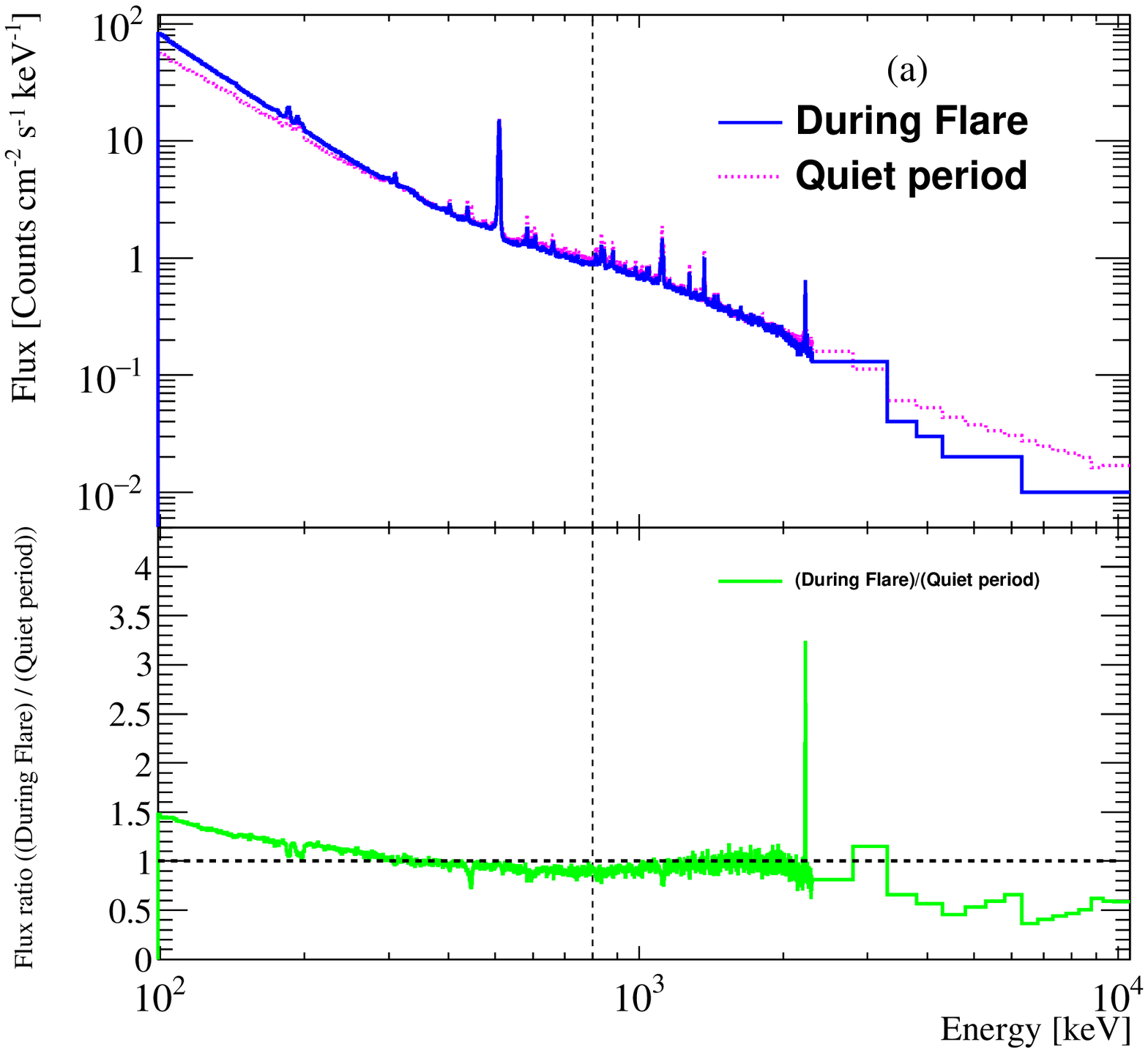}}
\end{minipage}
\begin{minipage}{0.49\hsize}
\centerline{\includegraphics[width=1.0\textwidth,clip=]{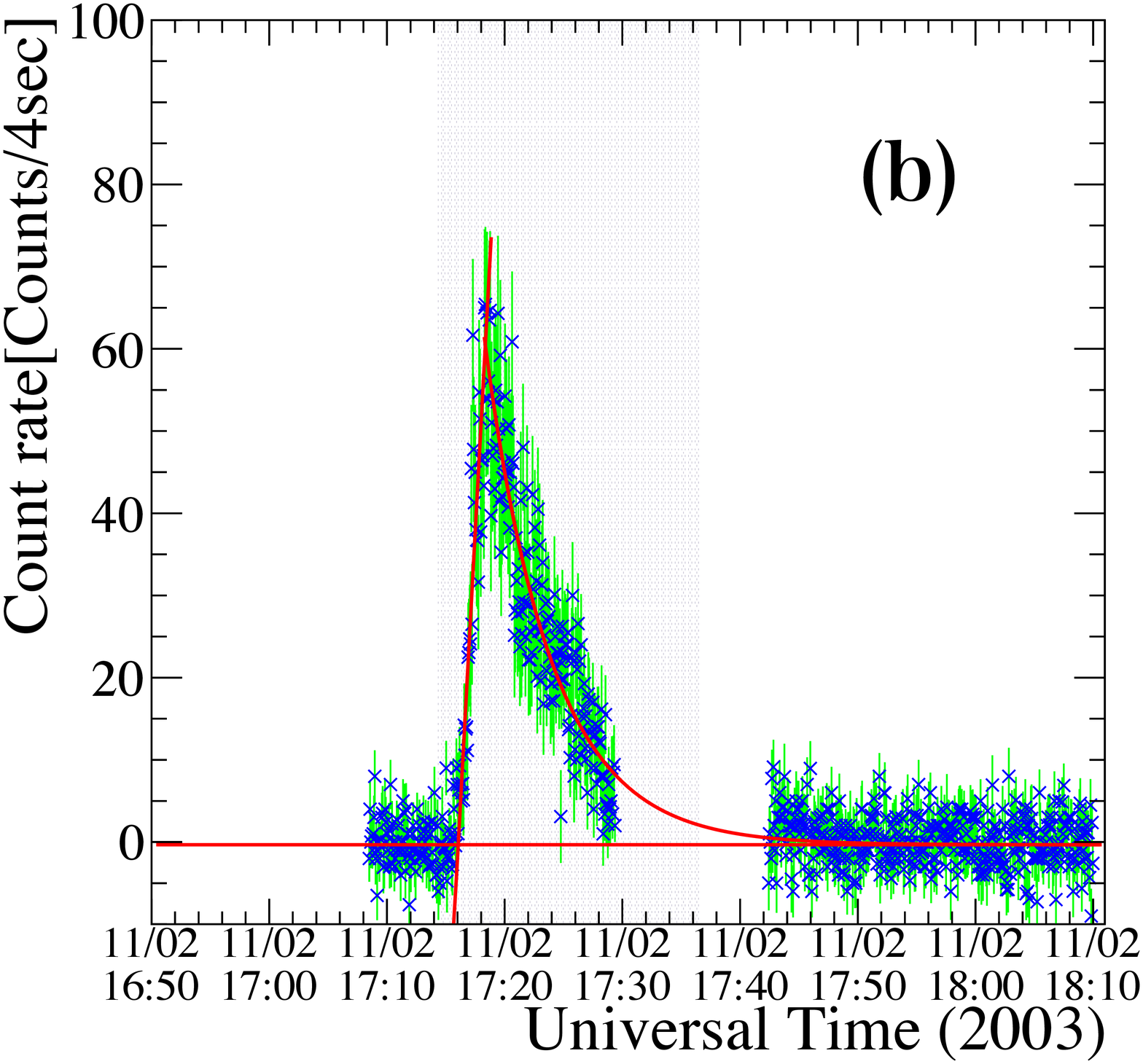}}
\end{minipage}
}
\caption{(a) Energy spectra of $\gamma$-rays associated with the solar flare occurring on November 2, 2003 as recorded by the RHESSI detectors (front 4--8 and rear 1--9). The blue solid and pink dotted lines shows the energy spectrum in quiet period and during solar flare~($17:03:00$--$18:07:03$ UTC). The horizontal axis shows energy in keV and the vertical axis in the top panel shows the flux of $\gamma$-rays in units of $\mathrm{cm^{-2} sec^{-1} keV^{-1}}$. The vertical axis in the bottom panel shows the ratio between fluxes.  The vertical dashed line in top and bottom panel indicates $800$~keV. For hard X-ray analysis, we used $100$--$800$~keV. Above this energy region, we searched for line $\gamma$-rays due to neutron captures. (b)~Light curve of the line $\gamma$-rays in this solar flare shown in (a). The horizontal axis shows universal time and the vertical axis shows the count rate in units of count per 4~s. The blue points and green bars depict the data and their error, respectively. The red lines show the fitting results and the shaded region shows the time window determined by this method.}\label{fig:rhessi}
\end{figure}

For line $\gamma$-ray analysis, we first scanned the energy spectrum recorded by the rear detector of RHESSI in quiet period of the Sun and during each solar flare as exampled in the top panel of Figure~\ref{fig:rhessi}a. During solar flare, we selected the time interval corresponding to the time window determined by GOES soft X-rays. On the other hand, we selected  the time interval of quiet period one month before a solar flare and then removed the time period when C (or larger) class flare occurred. Then, we took a ratio between them as shown in the bottom panel of Figure~\ref{fig:rhessi}a. When we confirmed an excess of the line $\gamma$-ray in the spectrum during the solar flare, we plotted the light curve using the energy range of $2.218$--$2.228$~MeV after subtracting its sideband of both $2.213$--$2.218$~keV and $2.228$--$2.233$~keV in order to remove bremsstrahlung from high energy electrons as shown in Figure~\ref{fig:rhessi}b. To determine the search window, we used the same method described above for hard X-rays. However, because of the neutron capture time delay~\citep{gan}, we ultimately selected the region of  [$t_{\mathrm{start}}-100~\mathrm{sec},~t_{\mathrm{end}}$] as the time window for line $\gamma$-ray.

\subsection{GEOTAIL Satellite Hard X-ray Events} \label{subsec:neutron}

The GEOTAIL satellite was launched in July 1992 to observe the Earth's magnetosphere. Its main targets are electrical field, magnetic field, plasma and high energy particles in the magnetosphere. The LEP~(low energy particle) instrument mounted on the satellite, can measure counts of ions and electrons independently based on energy-per-charge separation~\citep{mukai}. The LEP detector consists of an electrostatic analyzer and particle counters: microchannel plates for ions, and channel electron multipliers for electrons.

\begin{figure}
\centerline{\includegraphics[width=0.85\textwidth,clip=]{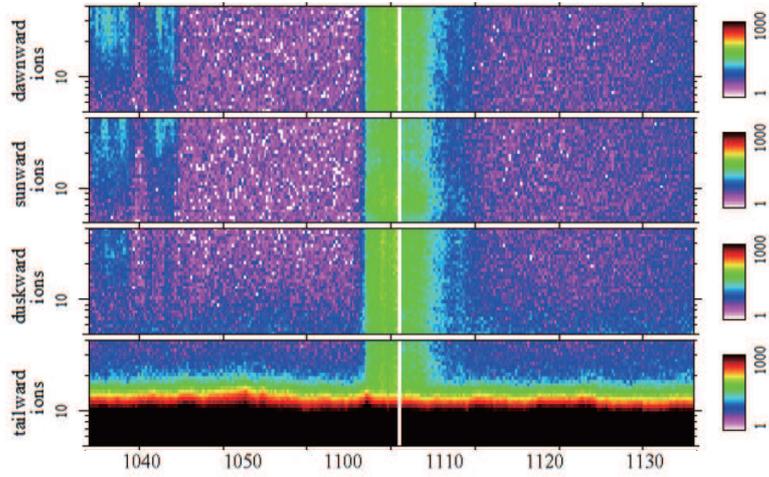}}
\caption{Count rates recorded by the LEP detector when this solar flare occurred on October, 28, 2003 are color-coded (white-blue-green-yellow-red-black). The horizontal axis shows the universal time (HH:MM) of particle detection. The vertical axis shows the energy of nonthermal ions in unis of energy-per-charge ($5$--$40~\mathrm{keV/q}$). Four panels, top to bottom, show ion counts in the four sectors pointed dawnward, sunward, duskward, and tailward,  respectively, where the ion flow directions are defined with respect to the relative magnetosphere geometry of the Sun and Earth.}\label{fig:geotail}
\end{figure}

GEOTAIL has occasionally been exposed to the intense fluxes of energetic ions ${>}50$~MeV and photons of ${>}50$~keV after strong solar flares or giant flares of magnetars~\citep{terasawa,geo3,geo2}. These energetic ions and photons do not follow the the electrostatic analyzer's energy-per-charge regulation, but rather penetrate through the satellite wall and hit the particle counters directly. For the solar flare case, energetic photons include hard X-ray photons from bremsstrahlung by high energy electrons, as well as line $\gamma$-ray photons from nuclear interactions. Figure~\ref{fig:geotail} shows an example of this situation. During the interval 10:35:00--11:35:00 UTC on October 28, 2003, GEOTAIL was in the solar wind upstream of the Earth's bow shock and continued to measure solar wind particles as well as nonthermal ions~($10$--$40$ $\mathrm{keV/q}$) accelerated by the bow shock. However, during the interval 11:01:40--11:15:30 UTC, GEOTAIL was hit the solar flare hard X-ray photons, which produced the energy-independent stripes in the first to three panels of Figure 3. It is noted that tailward-flowing solar wind ions~(less than ${\sim}20~\mathrm{keV/q}$ colored with yellow-red-black) overwrapped with the hard X-ray stripes as shown in the fourth panel of Figure 3. The penetration depth for hard X-ray photons from the satellite wall to the counter depends on the satellite spin phase, so that the colors of the energy-independent differ stripes slightly sector by sector. The electron detector (not shown) responded to the solar flare hard X-ray photons similarly. However, because the form factor of the electron detectors is two orders of magnitude smaller than that of the ion detectors, the apparent contribution of solar flare hard X-ray photons is much weaker.

\begin{figure}
\centerline{\includegraphics[width=0.8\textwidth,clip=]{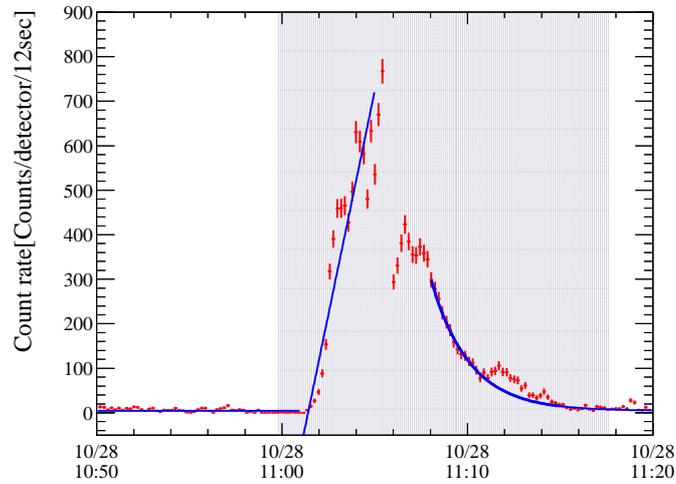}} 
\caption{Light curve associated with the solar flare occurring on October, 28, 2003 as recorded by the GEOTAIL satellite. The horizontal axis shows the universal time, and the vertical axis shows count rate in units of counts/detector/$12$~s). Red points show the data, and blue lines show the fitting results. This flare is the same flare shown in Figure 3.}\label{fig:geotail_fit}
\end{figure}

In the present study, we estimated the time profiles of particle acceleration by selecting data from the four sectors and the energy range that was not ``contaminated'' by ions, taking count averages every $12$~s. Figure~\ref{fig:geotail_fit} shows the photon counts from 10:50:00 to 11:20:00 UTC, as calculated from the data in the dawnward, sunward, and duskward sectors in Figure~\ref{fig:geotail}. Finally, the time window for GEOTAIL was determined using the same method as that used in the analysis for line $\gamma$-ray above, because a component of the GEOTAIL signal contains line $\gamma$-rays.

In order to avoid confusion with the term hard X-ray by RHESSI, we use the term hard X-ray~(${>}50$~keV, or above $50$~keV) to refer this channel by GEOTAIL even if the actual signals contain soft $\gamma$-ray as listed in Table~\ref{tbl:energy_range}.

%\clearpage
\section{Results and Discussion} \label{sec:result}

\subsection{Results} \label{subsec:result}

\begin{figure}
\centerline{\includegraphics[width=0.8\textwidth,clip=]{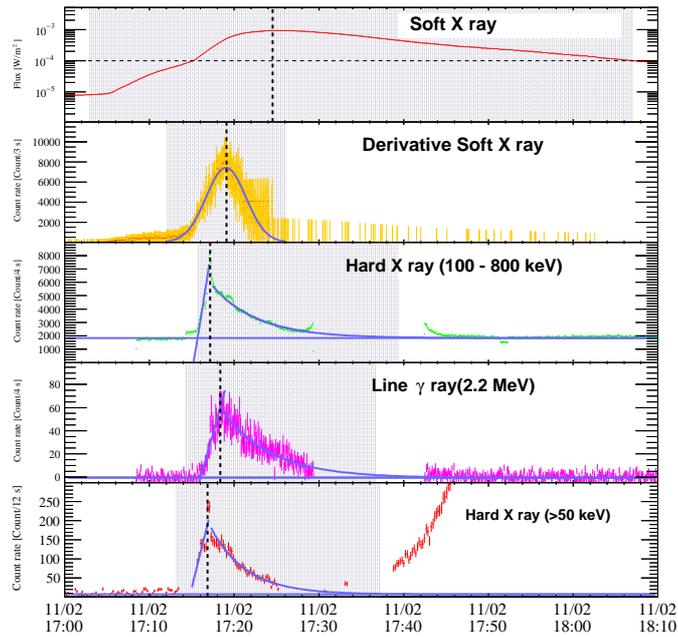}}
\caption{Light curves of the solar flare occurring on November 2, 2003. Gray regions show the time windows extracted by the analysis methods developed in this study. From top to bottom, the panels show the light curves for soft X-rays, derivative of soft X-rays, hard X-rays~($100$--$800$~keV), line $\gamma$-rays~($2.223$~MeV), and hard X-rays~(GEOTAIL, above $50$~keV), respectively. The vertical lines show the peak timing of each channel. For the hard X-ray light curve ~(above $50$~keV), the count rate gradually increased from 17:40 onward because protons in the range of $30$--$50$~MeV were observed. These protons masked other counts due to plasma particles and hard X-ray photons.}\label{fig:light_curves}
\end{figure}

Because all satellite data are available for the solar flare occurring on November 2, 2003, we used this example to illustrate the method for the time window determination developed in this study as shown in Figure~\ref{fig:light_curves}.

The analysis results for $23$ solar flares selected in this study are summarized in Tables~\ref{table:window1} and~\ref{table:window2}. It is noted that we failed to obtain the time window determined by derivative of soft X-ray for the solar flare occurred on September 9, 2005, because the brightening of its soft X-ray light curve was relatively slow and derivative of soft X-ray was not large enough. Except for this flare, we determined the time window for $22$  of $23$ solar flares by using derivative of the soft X-ray. In addition to this, we also determined the time window for five flares by hard X-ray~($100$--$800$~keV), three flares by line $\gamma$-ray, and seven flares by hard X-ray~(${>}50$~keV). It is noted that the time profile of the flare occurring on October 29, 2003 is not similar to those of the other flares, because a signal whose origin is not solar flare contaminated the solar flare signal. We did not determine the time windows for this flare by hard X--ray and line $\gamma$--ray channel. An alternative method, imaging method developed in \citealp{Liu_2009}, allows us to confirm non-solar flare signal causes this contamination. The hard X-ray light curve of this solar flare measured by this imaging method is shown in appendix. It is noted that there is not the line $\gamma$--ray light curve with the imaging method, because its flux was not large ehough. The line $\gamma$--ray time profile of this solar flare was investigated in \citealp{kurt2017}.

\begin{figure}
\centerline{\hspace*{0.015\textwidth}
\begin{minipage}{0.49\hsize}
\centerline{\includegraphics[width=1.0\textwidth,clip=]{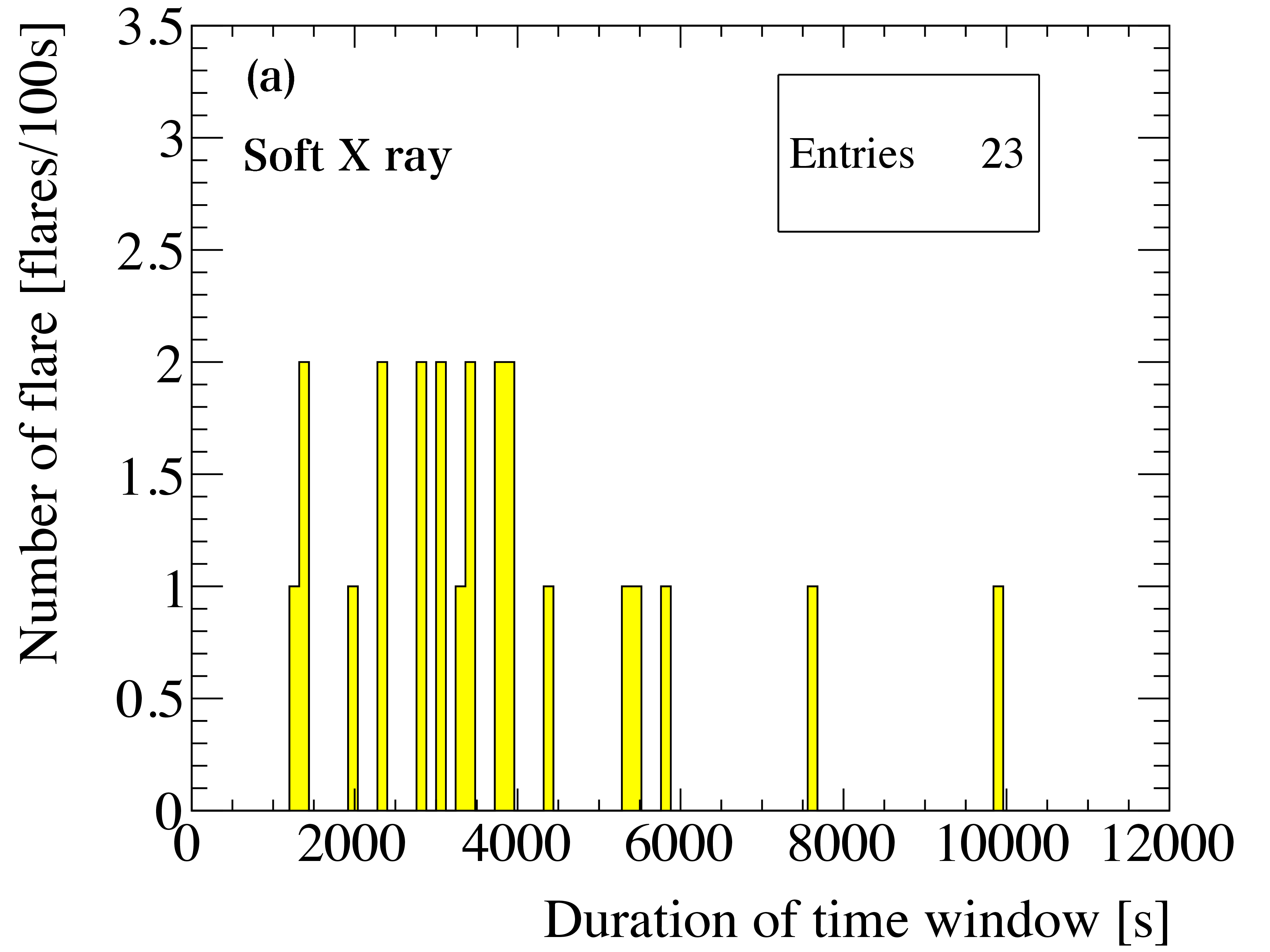}}
\end{minipage}
\begin{minipage}{0.49\hsize}
\centerline{\includegraphics[width=1.0\textwidth,clip=]{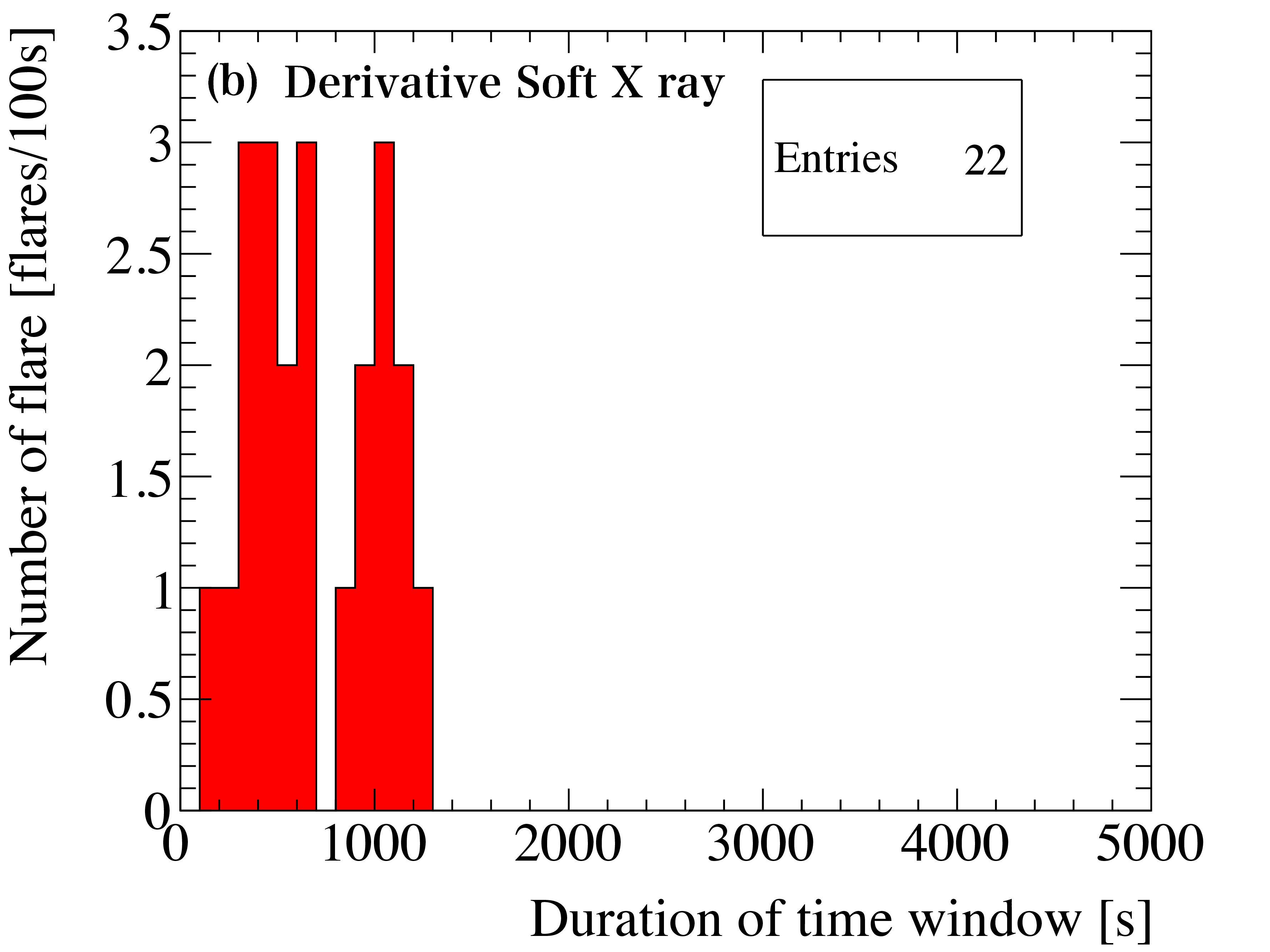}}
\end{minipage}
}
\centerline{\hspace*{0.015\textwidth}
\begin{minipage}{0.49\hsize}
\centerline{\includegraphics[width=1.0\textwidth,clip=]{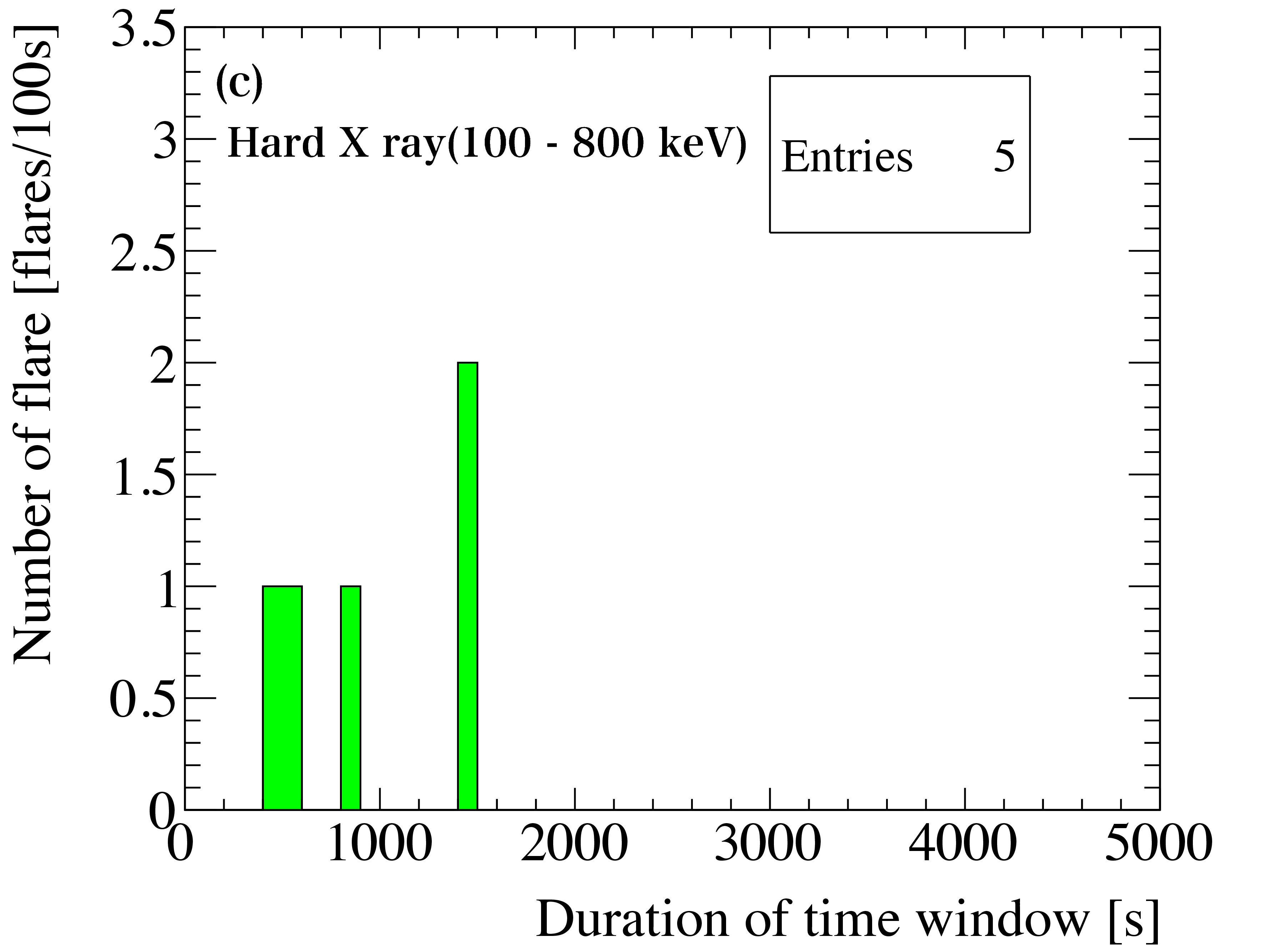}}
\end{minipage}
\begin{minipage}{0.49\hsize}
\centerline{\includegraphics[width=1.0\textwidth,clip=]{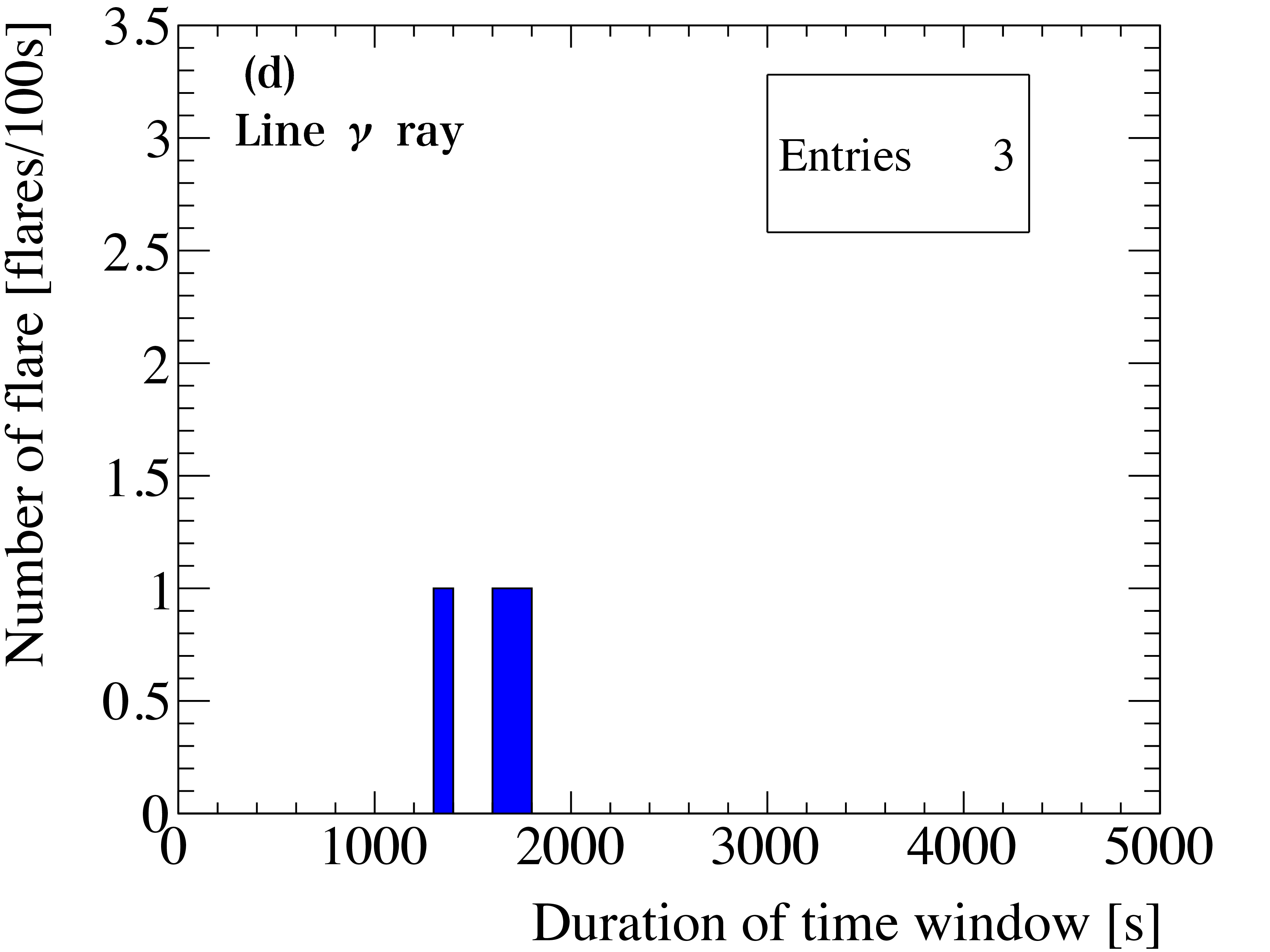}}
\end{minipage}
}
\centerline{\hspace*{0.015\textwidth}
\begin{minipage}{0.49\hsize}
\centerline{\includegraphics[width=1.0\textwidth,clip=]{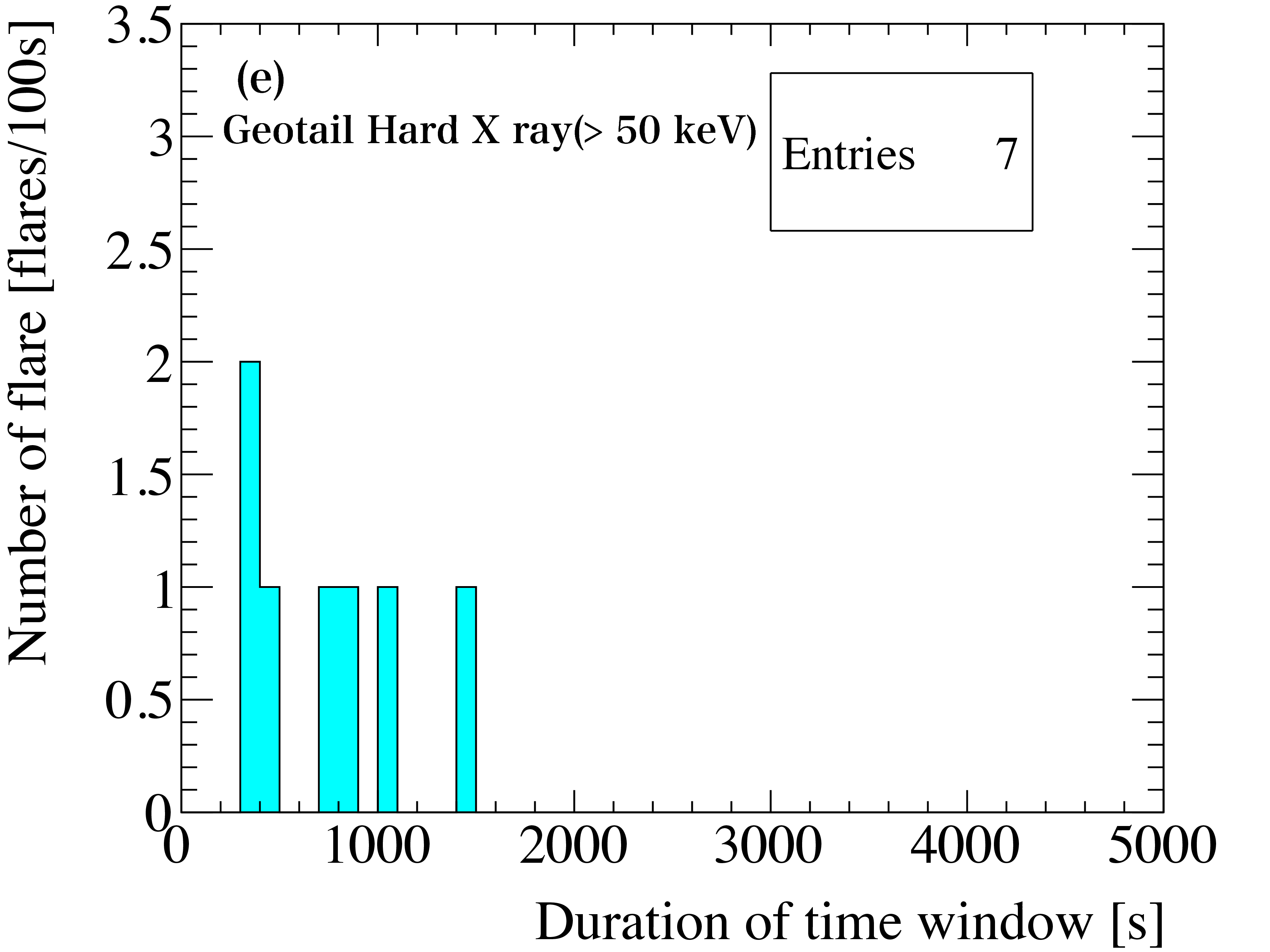}}
\end{minipage}
}
\caption{Distributions of the time window determined by each method: (a)~light curve of soft X-rays, (b)~derivative of the light curve of soft X-rays, (c)~hard X-rays~($100$--$800$~keV), (d)~line $\gamma$-rays, and (e)~hard X-rays~(${>}50$~keV). The horizontal axis shows duration of the time window and the vertical axis shows the number of entry.}
\label{fig:time_dist}
\end{figure}

Figure~\ref{fig:time_dist} summarizes the distribution of timing window duration for each channel. On the basis of these results, the average time duration for each channel is as follows; $4,178$~s for soft X-rays, $700$~s for derivative of soft X-rays, $944$~s for hard X-rays~($100$--$800$~keV), $1,586$~s for line $\gamma$-rays from neutron capture, and $776$~s for hard X-rays~(${>}50$~keV). The shortest time window among these is the derivative of soft X-rays.

\begin{landscape}
\begin{table}
%\scalebox{0.65}{
%  \begin{center}
    \caption{Time windows determined for soft X-rays, derivative of soft X-rays and hard X-rays. The time is described in units of UTC ([HH:MM:SS]).} \label{table:window1}
    \begin{tabular}{c|ccc|ccc|ccc} \hline
      Date&\multicolumn{3}{c|}{Soft X-rays~(GOES)}&\multicolumn{3}{c|}{Derivative of soft X-rays~(GOES)}&\multicolumn{3}{c}{Hard X-rays~(RHESSI)}\\ 
       & $t_{\rm{start}}$&$t_{\rm{end}}$&$\Delta t$~[s]&$t_{\rm{start}}$&$t_{\rm{end}}$&$\Delta t$~[s]&$t_{\rm{start}}$&$t_{\rm{end}}$&$\Delta t$~[s]\\ \hline
      1997 Nov. 6 & 11:49:00 & 12:12:21 &1,401 & 11:52:13&11:54:58&165&--&--&--\\ 
      2000 Jul. 14& 10:03:00 & 11:05:03 &3,723 & 10:08:44&10:28:31&1,187&--&--&--\\ 
      2001 Apr. 2& 21:32:00 & 22:56:12 &5,052 & 21:35:10&21:52:57&1,067&--&--&--\\ 
      2001 Apr. 6& 19:10:00 & 19:43:15 &1,995 & 19:12:32&19:21:31&539&--&--&-- \\ 
      2001 Apr. 15& 13:19:00 & 14:23:39 &3,879 & 13:43:48&13:51:03 &435&--&--&-- \\ 
      2001 Aug. 25& 16:23:00 & 17:36:18 &4,398 & 16:28:36&16:33:38&302&--&--&--\\ 
      2001 Dec. 13& 14:20:00 & 14:41:57 &1,317 & 14:24:25&14:31:05&400&--&--&-- \\ 
      2002 Jul. 23& 00:18:00 & 01:08:27 &3,027 & 00:25:08&00:31:36&388&00:25:44&00:36:46&562 \\ 
      2003 Oct. 23& 08:19:00 & 09:13:9 &3,249 & 08:20:32&08:36:03&931&--&--&-- \\ 
      2003 Oct. 28& 09:51:00 & 12:36:54 &9,954 & 10:59:30&11:10:57&687&--&--&-- \\ 
      2003 Oct. 29& 20:37:00 & 21:23:39 &2,799 & 20:38:46&20:50:07&681&--&--&-- \\ 
      2003 Nov. 2& 17:03:00 & 18:07:03 &3,843 & 17:12:01&17:26:03&842&17:15:41&17:39:22&1421 \\ 
      2003 Nov. 4& 19:29:00 & 21:10:12 &6,072 & 19:36:59&19:56:03&1,144&--&--&-- \\ 
      2005 Jan. 20& 06:36:00 & 08:05:48 &5,388 & 06:39:22&06:57:12&1,070&06:43:27&06:57:57&870 \\ 
      2005 Sep. 7& 17:17:00 & 18:48:57 &5,517 & 17:23:34&17:39:48&974&--&--&-- \\ 
      2005 Sep. 8& 20:52:00 & 21:31:48 &2,388 & 21:00:27&21:08:05&458&--&--&-- \\ 
      2005 Sep. 9& 19:13:00 & 21:20:36 &7,656 & --&--&--&--&--&--\\ 
      2006 Dec. 5& 10:18:00 & 10:56:09 &2,289 & 10:24:18&10:35:43&685&--&--&-- \\ 
      2006 Dec. 6& 18:29:00 & 19:16:30 &2,850 & 18:41:29&18:47:06&337&18:42:33&18:49:19&406 \\ 
      2011 Aug. 9& 07:48:00 & 08:11:47 &1,427& 08:00:50&08:05:40&290&--&--&-- \\ 
      2012 Mar. 7& 00:02:00 & 01:31:34 &5,374 & 00:04:21&00:25:16&1,255&--&--&--\\ 
      2017 Sep. 6& 11:53:00 & 13:30:00 &5,820 & 11:54:39&12:03:20&521&--&--&-- \\
      2017 Sep. 10& 15:35:00 & 17:26:06 &6,666 & 15:49:12&16:06:32&1,040&15:54:42&16:19:03&1,461\\ \hline
    \end{tabular}
%} %scalebox
%  \end{center}
\end{table}
\end{landscape}

%\begin{landscape}
\begin{table}[h]
%  \begin{center}
    \caption{Time windows determined for line $\gamma$-rays~(RHESSI) and hard X-rays~(GEOTAIL). Time is described in units of UTC ([HH:MM:SS]).} \label{table:window2}
%    \scalebox{0.7}{
    \begin{tabular}{c|ccc|ccc} \hline
      Date&\multicolumn{3}{c|}{Line $\gamma$-ray~(RHESSI)}&\multicolumn{3}{c}{Hard X-ray~(GEOTAIL)} \\
       &$t_{\rm{start}}$&$t_{\rm{end}}$&$\Delta t$~[s]&$t_{\rm{start}}$&$t_{\rm{end}}$&$\Delta t$~[s]\\ \hline
      1997 Nov. 6&--&--&--&11:50:55&11:57:04&369\\
      2000 Jul. 14&--&--&--&--&--&--\\
      2001 Apr. 2&--&--&--&--&--&--\\
      2001 Apr. 6&--&--&--&--&--&-- \\
      2001 Apr. 15&--&--&--&13:43:42&13:51:58&469 \\
      2001 Aug. 25&--&--&--&16:28:29&16:34:54&385\\
      2001 Dec. 13&--&--&--&--&--&-- \\
      2002 Jul. 23&00:26:28&00:56:22&1,794&--&--&-- \\
      2003 Oct. 23&--&--&--&--&--&-- \\
      2003 Oct. 28&--&--&--&10:59:45&11:17:36&1,071 \\
      2003 Oct. 29&--&--&--&--&--&-- \\
      2003 Nov. 2&17:14:14&17:36:41&1,347&17:13:08&17:37:11&1,443 \\
      2003 Nov. 4&--&--&--&19:38:35&19:53:32&897 \\ 
      2005 Jan. 20&06:42:54&07:09:52&1,618&--&--&-- \\ 
      2005 Sep. 7&--&--&--&17:32:44&17:45:32&7678\\ 
      2005 Sep. 8&--&--&--&--&--&-- \\
      2005 Sep. 9&--&--&--&--&--&-- \\
      2006 Dec. 5&--&--&--&--&--&-- \\
      2006 Dec. 6&--&--&--&--&--&-- \\
      2011 Aug. 9&--&--&--&--&--&-- \\
      2012 Mar. 7&--&--&--&--&--&-- \\
      2017 Sep. 6&--&--&--&--&--&-- \\
      2017 Sep. 10&--&--&--&--&--&-- \\ \hline
    \end{tabular}
%	}%scale box
%  \end{center}
\end{table}
%\end{landscape}

\begin{figure}
\centerline{\includegraphics[width=0.8\textwidth,clip=]{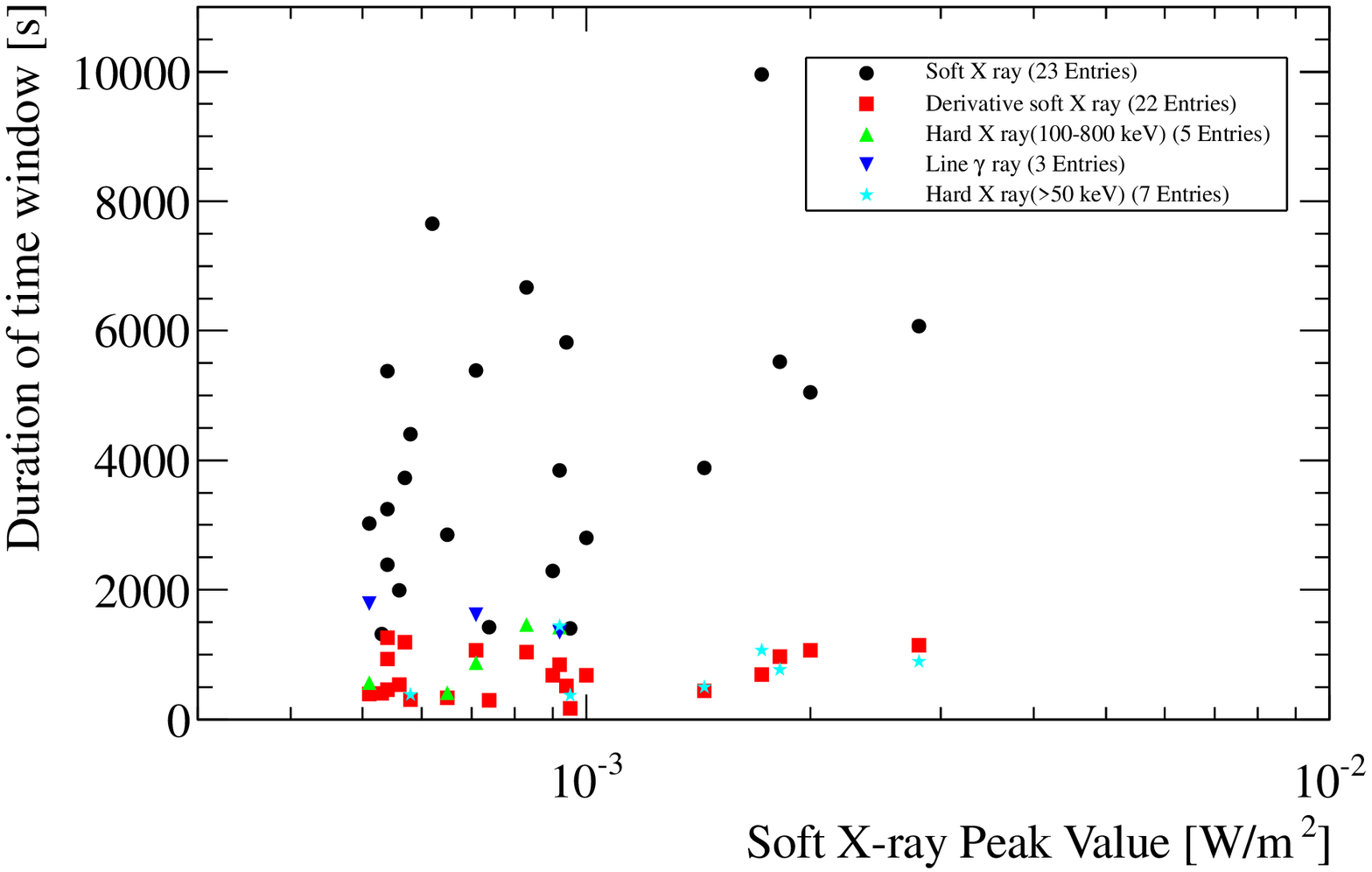}}
\caption{Time duration as a function of the peak intensity of soft X-ray. Black dot, red square, green up-triangle, blue down-triangle, and light blue star show soft X-ray, derivative of soft X-ray, hard X-ray~($100$--$800$~keV), line $\gamma$-ray, and hard X-ray~(${>}50$~keV), respectively.} \label{fig:time_range}
\end{figure}

Figure~\ref{fig:time_range} shows the time duration of search windows as a function of the maximum value of the soft X-ray peak. The correlation coefficients between the maximum value of the soft X-ray peak and the time duration of search windows are $0.436$ for soft X-rays, $0.341$ for soft X-rays~(derivative), $0.236$ for hard X-rays~(100--800~keV), $-0.344$ for line $\gamma$-rays, and $0.307$ for hard X-rays~(${>}50$~keV). Therefore, no strong correlation is found among them, as summarized in Table~\ref{tbl:comp_time}.

\begin{table}
\caption{Summary of timing information obtained in this study. Average time durations are calculated from the distributions shown in Figure~\ref{fig:time_dist}. Correlation coefficients are calculated from the distribution shown in Figure~\ref{fig:time_range}. The differences in peak timing are calculated at a soft X-ray peak time of $0$~(Figure~\ref{fig:diff_peak}).} \label{tbl:comp_time}
\begin{tabular}{ccccc}
\hline
 & Entries &Average duration & Correlation  & Peak timing \\
 & & [s] & coefficient & difference [s] \\ \hline
Soft X-ray &$23$ &  $4,178$ & $0.436$ & -- \\
Soft X-ray~(derivative) &$22$& $700$ & $0.341$ & $-320$ \\
Hard X-ray~($100$--$800$~keV) & $5$ & $944$ & $0.236$ & $-482$ \\
Line $\gamma$-ray &$3$& $1,586$ & $-0.344$ & $-471$ \\
Hard X-ray~(${>}50$~keV) &$7$& $776$ & $0.307$ &  $-342$ \\ \hline
\end{tabular}
\end{table}

\begin{figure}
\centerline{\includegraphics[width=0.8\textwidth,clip=]{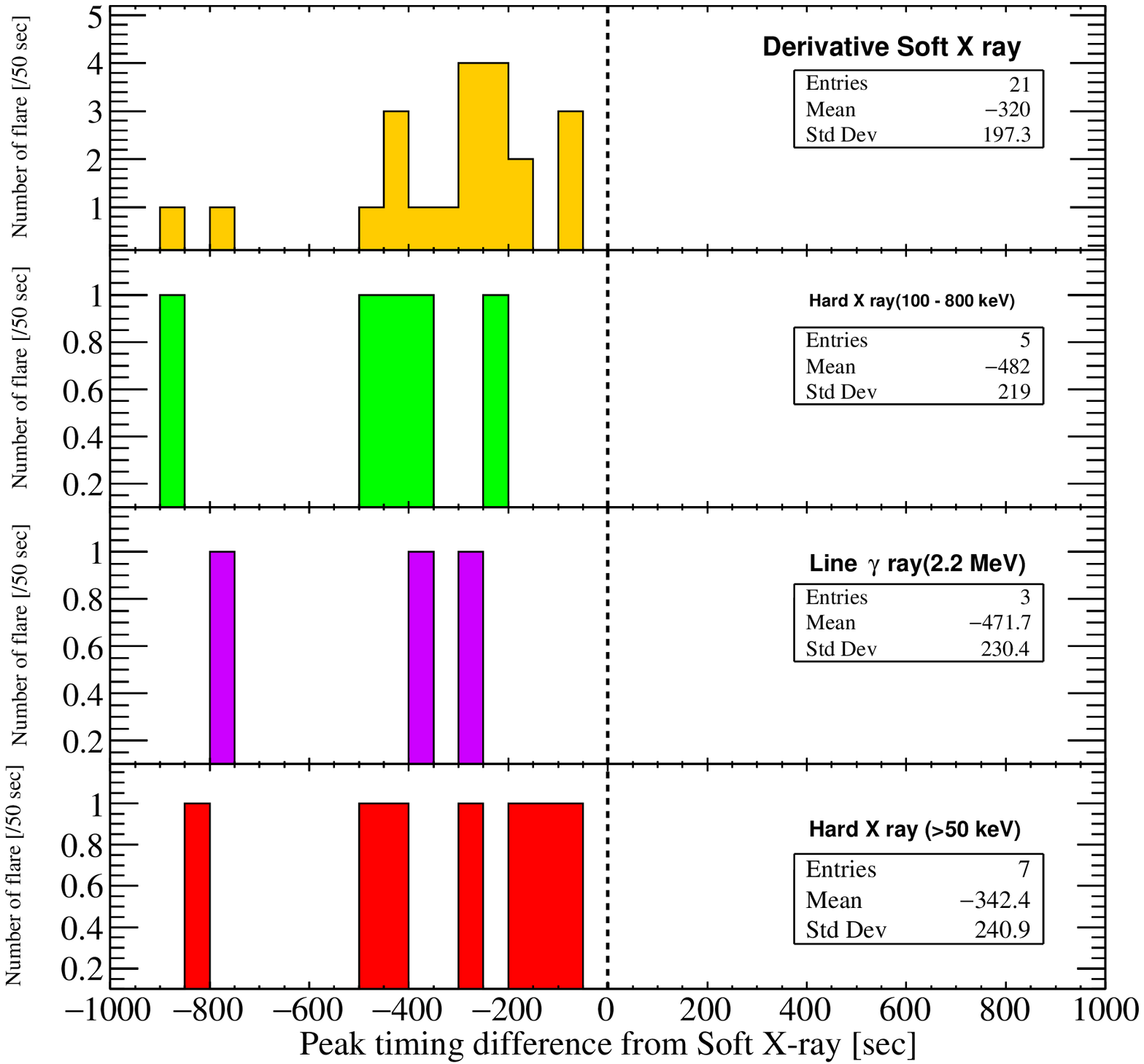}}
\caption{Time difference distribution after subtracting the peak timing of soft X-rays. The horizontal axis shows the time difference in units of seconds and the vertical axis shows the entry per $50$~s.} \label{fig:diff_peak}
\end{figure}

To investigate the difference in time profile between channels, we calculated the time elapsed between the peak timing of a given channel and the peak timing of soft X-rays as shown in Figure~\ref{fig:diff_peak}, because the peak timing of each channel is earlier than that of soft X-rays. The mean differences are $-320$~s for the derivative of soft X-rays, $-482$~s for hard X-rays~($100$--$800$~keV), $-471$~s for line $\gamma$-rays, and $-342$~s for hard X-rays~(${>}50$~keV).

\begin{figure}
\centerline{\hspace*{0.015\textwidth}
\begin{minipage}{0.49\hsize}
\centerline{\includegraphics[width=1.0\textwidth,clip=]{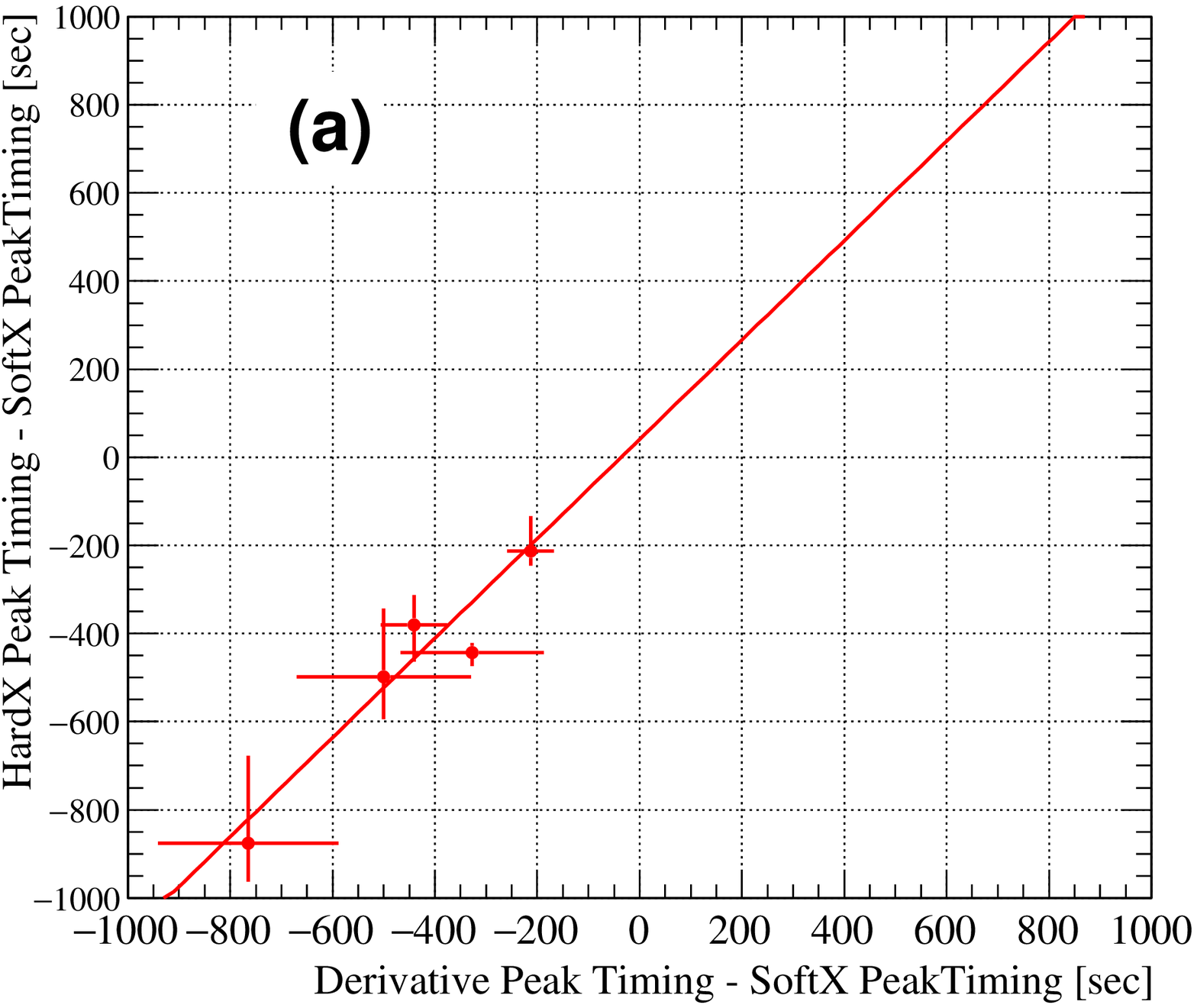}}
\end{minipage}
\begin{minipage}{0.49\hsize}
\centerline{\includegraphics[width=1.0\textwidth,clip=]{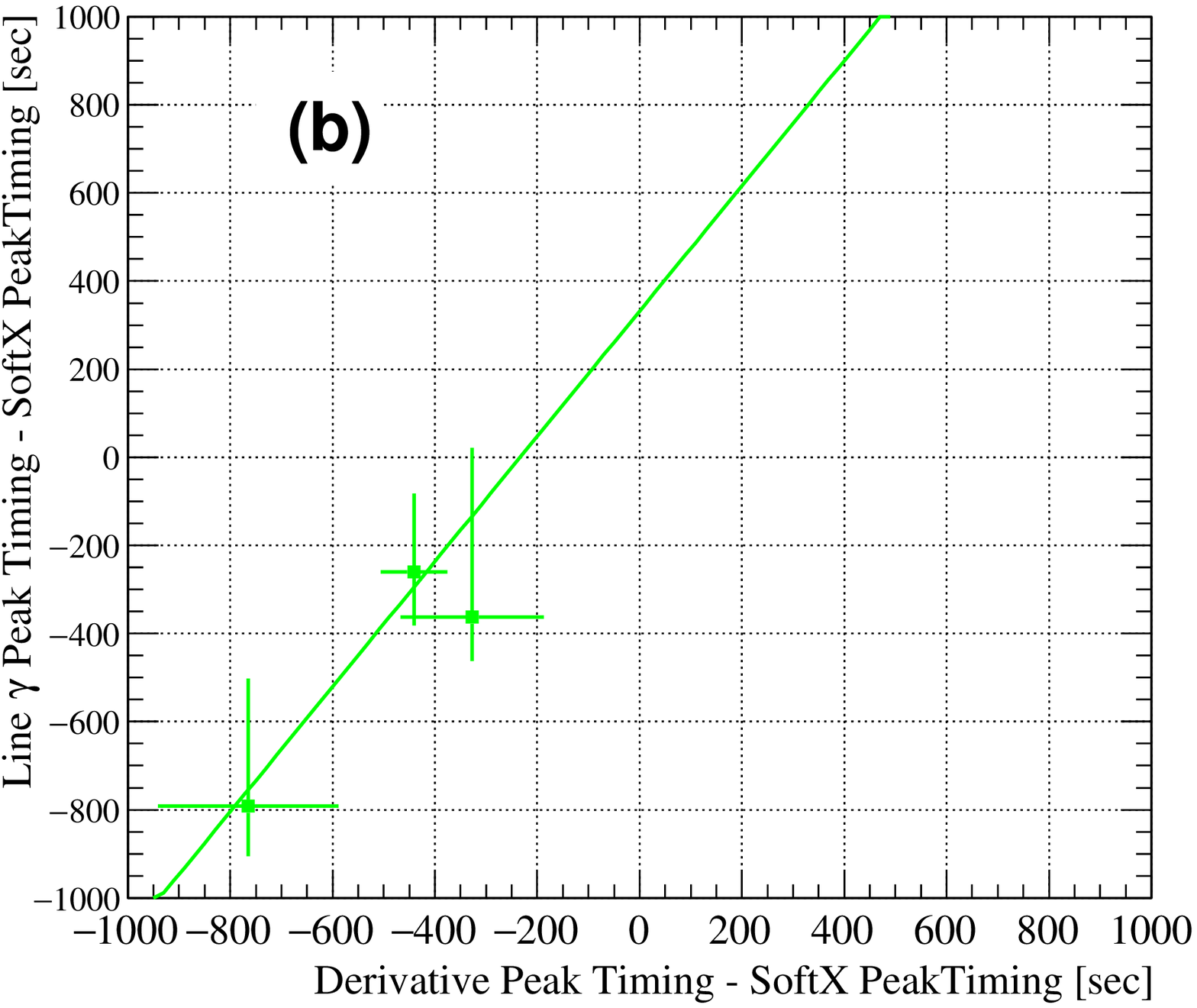}}
\end{minipage}
}
\centerline{\hspace*{0.015\textwidth}
\begin{minipage}{0.49\hsize}
\centerline{\includegraphics[width=1.0\textwidth,clip=]{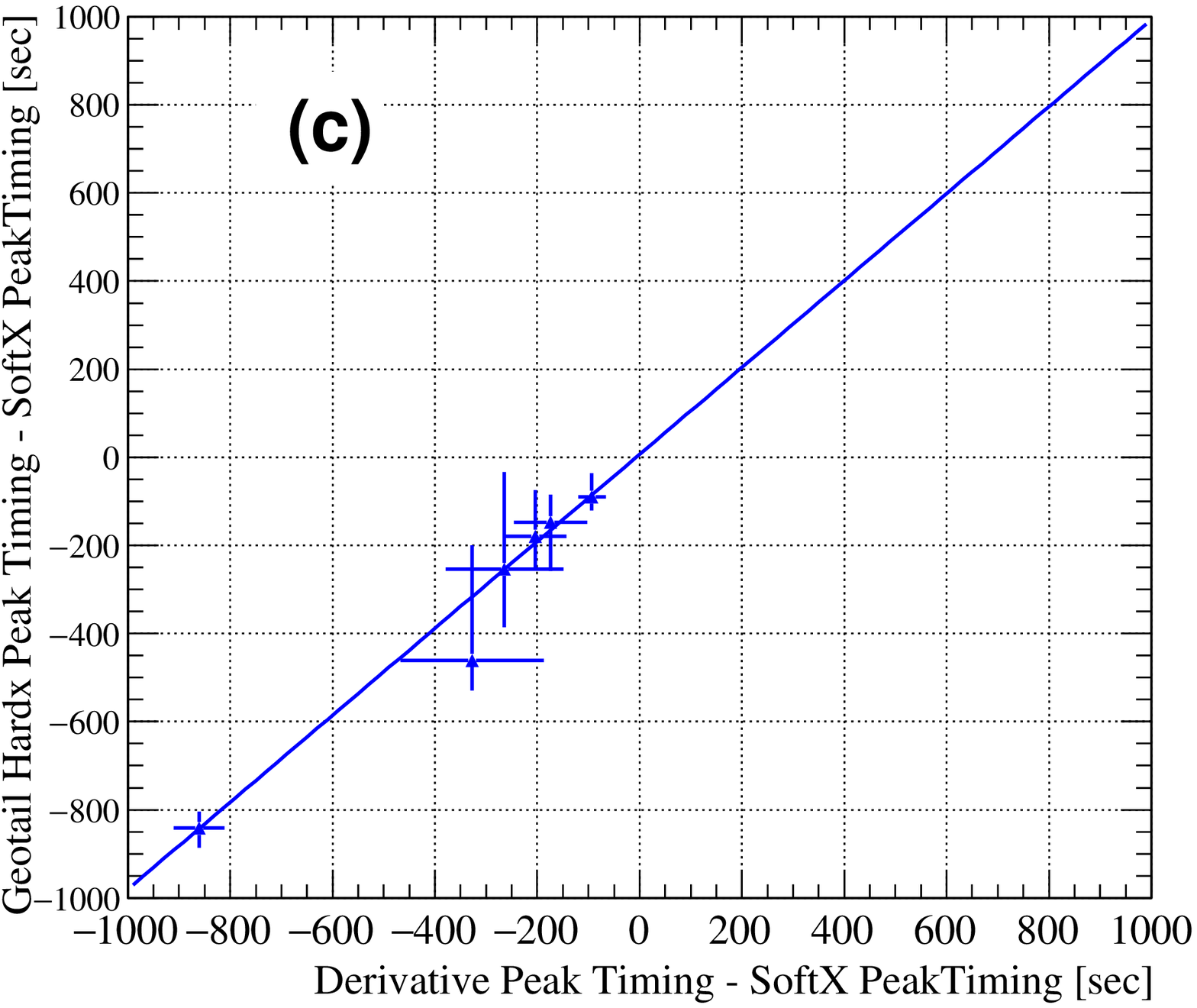}}
\end{minipage}
\begin{minipage}{0.49\hsize}
\centerline{\includegraphics[width=1.0\textwidth,clip=]{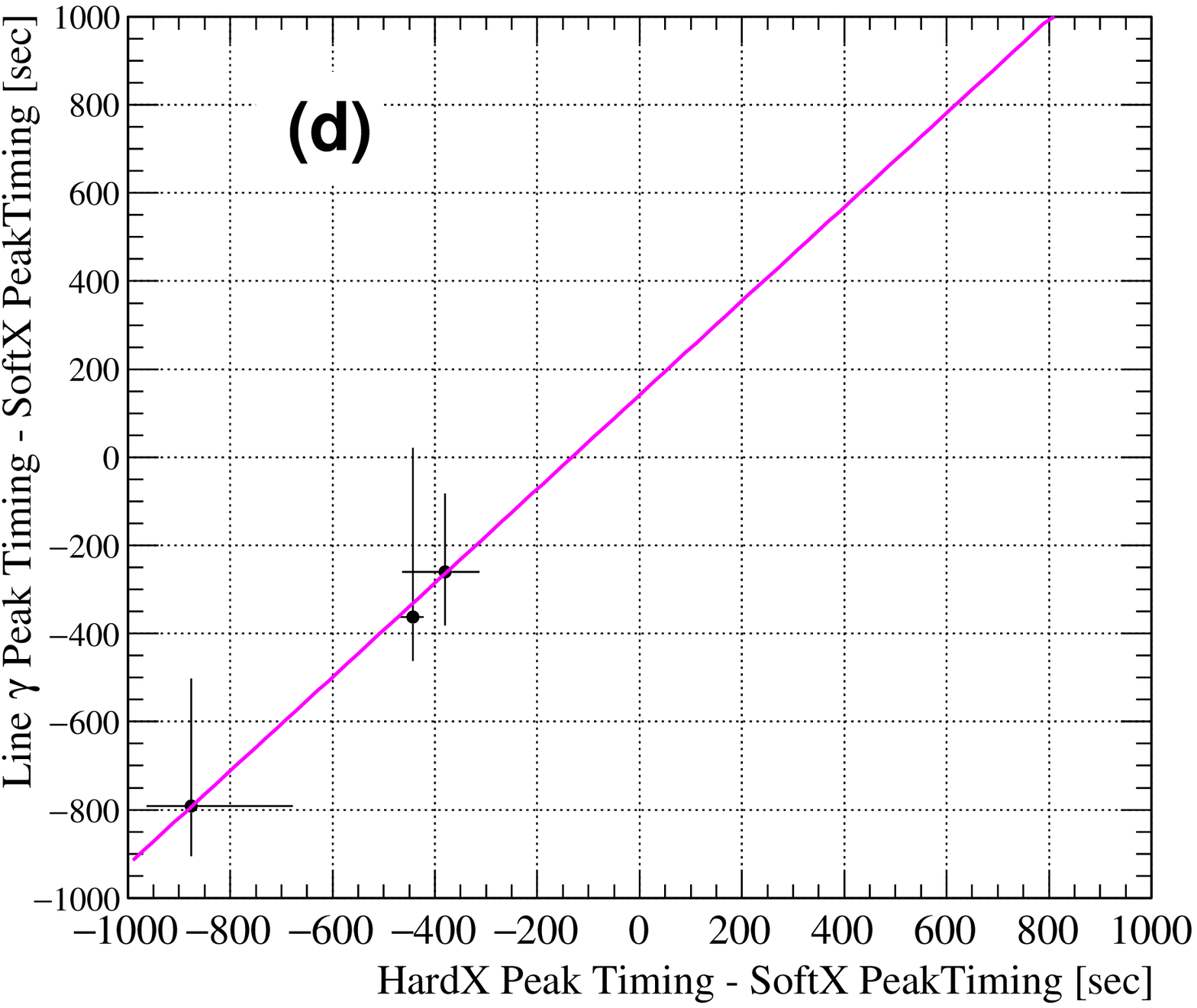}}
\end{minipage}
}
\centerline{\hspace*{0.015\textwidth}
\begin{minipage}{0.49\hsize}
\centerline{\includegraphics[width=1.0\textwidth,clip=]{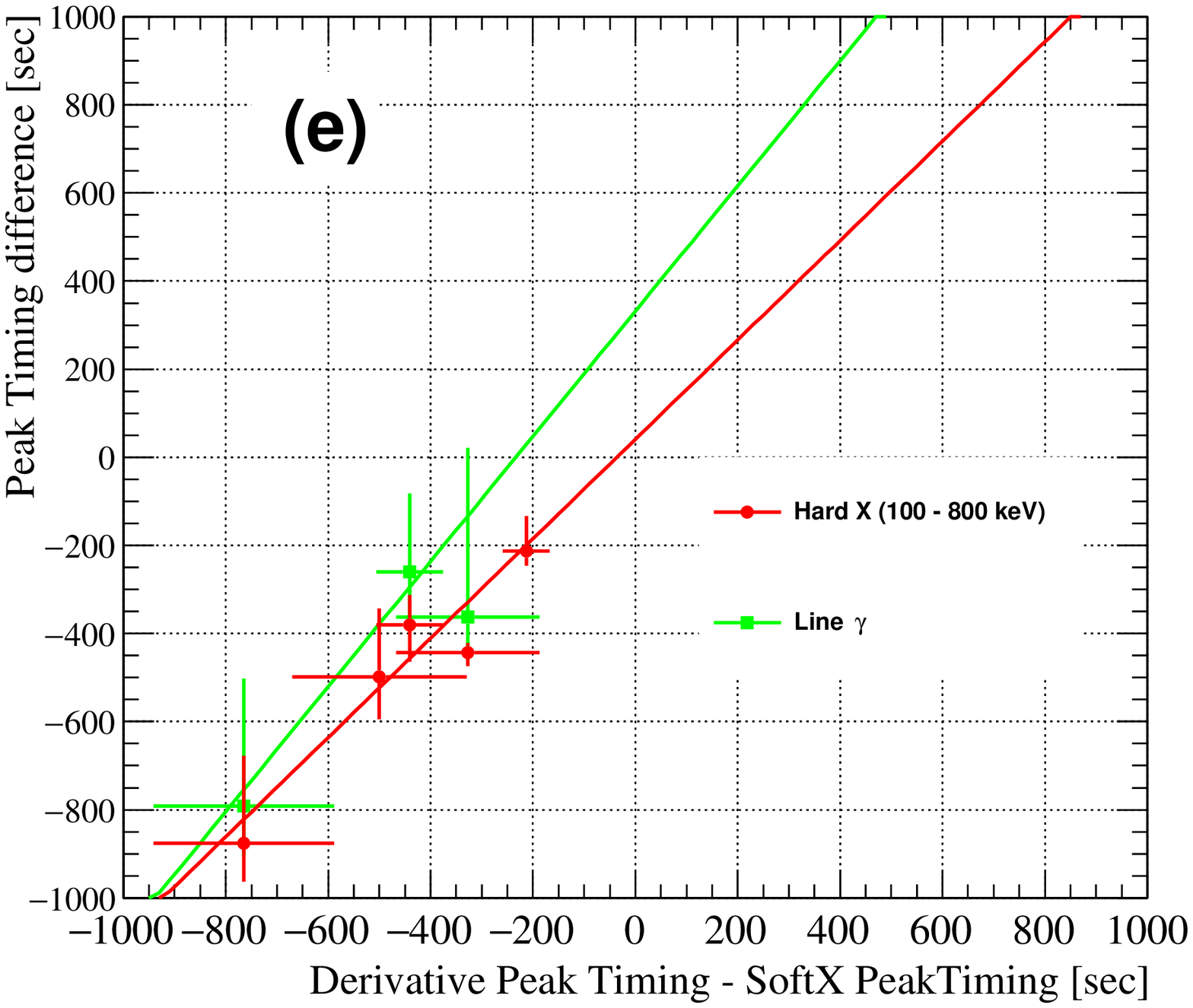}}
\end{minipage}
}
\caption{Relationship between each channel's peak timing after subtracting the peak time of soft X-rays. The fitting results are summarized in Table~\ref{tbl:summary_fit}. For (e), the plots shown in both (a) and (b) are overlaid. }
\label{fig:time_peak_diff}
\end{figure}

Each panel in Figure~\ref{fig:time_peak_diff} shows the time correlation between two variables: on the x-axis, the peak timing of soft X-rays; on the y-axis, a given channel after subtracting the peak timing of soft X-rays. To extract the difference between peak timings, the plots are fitted with a linear function; the fitting results are summarized in Table~\ref{tbl:summary_fit}. The y-intercept of the linear function gives the difference in peak timings.

In Figure~\ref{fig:time_peak_diff}c, the fitting results are $0.9\pm0.1$ for the slope and $7\pm55$~s for the y-intercept. We found a good timing correlation between the derivative of soft X-rays and hard X-rays because the y-intercept is close to the origin of coordinates~($0,0$), thus confirming the Neupert effect between them. Notably, this confirmation takes place by comparing the peak timing without considering the shape of light curves.

In Figure~\ref{fig:time_peak_diff}d, the y-intercept is $141\pm285$~s, suggesting that the peak timing of line $\gamma$-rays is delayed from that of hard X-rays~($100$--$800$~keV). If the acceleration of both electrons and ions occurs simultaneously, this difference between their timings is explained by considering neutron capture~\citep{gan,hard-line}.

Figure~\ref{fig:time_peak_diff}e overlays the plots shown in Figure~\ref{fig:time_peak_diff}a and Figure~\ref{fig:time_peak_diff}b. We found that the peak timing of hard X-rays is earlier than that of line $\gamma$-rays, where their difference is about $100$~s.

\begin{table}
\caption{Summary of results from fits in Figure~\ref{fig:time_peak_diff}. Each axis in Figure~\ref{fig:time_peak_diff} represents the time elapsed between the peak timing of a given channel and that of soft X-rays.} \label{tbl:summary_fit}
\begin{tabular}{ccccc}
\hline
 Figure~\ref{fig:time_peak_diff} & x-axis & y-axis  & Fitting result & $\chi^{2}/\mathrm{d.o.f}$ \\ \hline
(a) & Soft X-ray& Hard X-ray & $(1.1\pm0.4)x$ & $1.0/3$ \\
 & (derivative)  & ($100$--$800$~keV) & $+(41\pm166)$ &  \\ \hline
(b) & Soft X-ray & Line $\gamma$-ray & $(1.4\pm1.1)x$ & $0.3/1$ \\
 & (derivative)  & & $+(331\pm548)$ &  \\ \hline
(c) & Soft X-ray & Hard X-ray & $(0.9\pm0.1)x$ & $0.3/4$ \\
 & (derivative)  & (${>}50$~keV) & $+(7\pm55)$ &  \\ \hline
(d) & Hard X-ray& Line $\gamma$-ray & $(1.0\pm0.5)x$ & $0.1/1$ \\
 & ($100$--$800$~keV)   & & $(141\pm286)$ &  \\ \hline
\end{tabular}
\end{table}

\subsection{Applications to Neutrino Detectors}

On the basis of our results, the longest time window, averaging $4,178$~s, can be set using soft X-rays. It is expected that all processes (acceleration of charged particles, energy release of (non)~thermal electrons, etc) occurring during a solar flare complete within this time window. Therefore, this window is advantageous for estimating the time scale of physics processes during solar flares. Despite its uselessness for identifying the time of solar flare neutrinos, the continuous monitoring by GOES allows us to set the search window for every solar flare even if other satellites miss the signals.

The time derivative of soft X-rays allows the setting of the shortest time window, with an  average of $700$~s. This time window is advantageous for searching for solar flare neutrinos, because it has the best chance of improving the signal-to-noise ratio. In the case of Super-K, the average observing rate of atmospheric neutrinos is $8.3$~events/day~\citep{atm_sk}. Assuming this time window, the expected background rate can be reduced to $0.067$~events/flare. If three events are observed within this time window, the Super-K detector can reject the null hypothesis of neutrinos not from solar flares at $95\%$ confidence level (C.L.). On the contrary, the derivative of soft X-rays does not provide us with information on the acceleration of  protons or ionized particles, which produces neutrinos via hadronic interactions. Therefore, it is questionable whether this window is entirely appropriate for a comprehensive solar flare neutrino search.

The line $\gamma$-ray observation is crucial in this study because such signals indicate the acceleration of protons and hadronic interactions including charged pions that produce neutrinos. The search window for solar flare neutrinos should therefore be determined by this channel. In the present study, four solar flares include $\gamma$-ray observation by RHESSI. One of these is shown in Figure~\ref{fig:light_curves}. The light curves of the other three flares are shown in Figures~\ref{fig:light_10}--\ref{fig:light_12} in the appendix. According to our study, hard X-ray~($100$--$800$~keV) light curve shape looks similar to that of line $\gamma$-rays. Their time profiles are also similar after subtracting the peak timing of soft X-rays in Figure~\ref{fig:time_peak_diff}c. Therefore, we suggest that the search window determined by hard X-ray~($100$--$800$~keV) information is useful in the  search for solar flare neutrinos even if no line $\gamma$-ray data are available.

\section{Conclusions and Future Prospects}

Solar flare neutrinos attract significant attention because the detection of such neutrinos can extract information about particle acceleration during solar flares. Hence, we have developed a method for determining the search window for solar flare neutrinos. The longest search window, based on soft X-ray light curve, is estimated to last for $4,178$~s. On the contrary, the shortest search window, based on the derivative of soft X-rays, is estimated to last for $700$~s.

The method developed in this study is useful for adoption by other space satellites, such as HINOTORI, Yohkoh, Hinode~\citep{hinode}, and several future planned satellites, for example the {\it Spectrometer/Telescope for Imaging X-rays} on board the {\it ESA Solar Orbiter}~\citep{stix,stix2}, the {\it Focusing Optics X-ray Solar Imager}~\citep{foxsi}, and {\it Physics of Energetic and Nonthermal plasmas in the X region}~(PhoENiX), proposed in Japan. This adoption will permit the extraction of  information about both the neutrino production timing and particle acceleration during solar flares.

Using the method in this study, a search for solar flare neutrinos above the X5 class undertaken from 1996 to 2018 by neutrino detectors, such as Super-K, SNO~\citep{sno}, IceCube, ANTARES~\citep{antares}, KamLAND~\citep{kamland} and Borexino~\citep{bore}. Assuming the shortest time window of derivative of soft X-ray, the expected background rate can be reduced to $0.067$~events/flare in the case of Super-K. If three events are observed within this time window, the Super-K detector can reject the null hypothesis of neutrinos not from solar flares at $95\%$~C.L.

This method can also be used for future large solar flares and can improve sensitivity in the next generation of neutrino experiments, such as Hyper-Kamiokande,  IceCube-Gen2~\citep{icegen2}, SNO${+}$~\citep{sno_p}, DUNE~\citep{dune}, LENA~\citep{lena}, and JUNO~\citep{juno}.


%% Figure
%
% \begin{figure}
% \centerline{\includegraphics[width=0.5\textwidth,clip=]{<fig.eps>}}
% \caption{}%\label{fig:?}
% \end{figure}

%% Table
%
% \begin{table}
% \caption{}%\label{tbl:?}
% \hline
% \multicolumn{2}{c}{<>}
% <data>
% \hline
% \end{tabular}
% \end{table}

%%%%%%%%%%%%%%%%%%%%%%%%%%%%%%%%%%%%%%%%%%%%%%%%%%%%%%%%%%%%%%%%%%%%%%%%%%%
%% Appendix
%
% \appendix

%%%%%%%%%%%%%%%%%%%%%%%%%%%%%%%%%%%%%%%%%%%%%%%%%%%%%%%%%%%%%%%%%%%%%%%%%%%
%% Acknowledgements
%
\begin{acks}
%\acknowledgments

This work was carried out by the joint research program of the Institute for Space-Earth Environmental Research~(ISEE), Nagoya University. A part of this study was carried using the computational resources of the Center for Integrated Data Science, Institute for Space-Earth Environmental Research, Nagoya University, through the joint research program. We thank Y.~Saito from JAXA and I.~Shinohara from JAXA for reproducing GEOTAIL satellite data.  We thank S.~Krucker for his careful reading of this manuscript and for his insightful comments and suggestions. This work is supported by MEXT KAKENHI Grant Numbers 17K17880, 18H05536, 18J00049 and 19J21344.
\end{acks}

\section*{Disclosure of Potential Conflicts of Interest}
The authors declare that they have no conflicts of interest.

\appendix
\section{Line $\gamma$-ray Light Curves for Solar Flares}

\begin{figure}
\centerline{\includegraphics[width=0.8\textwidth,clip=]{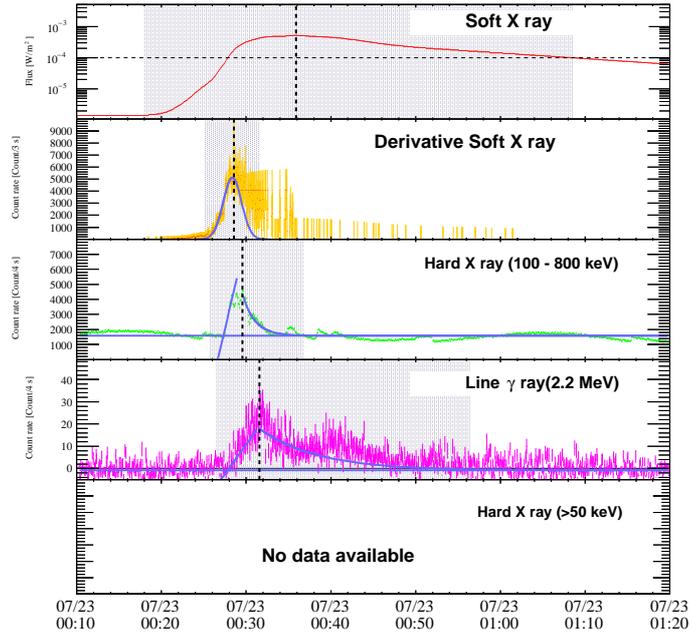}}
\caption{Light curves of the solar flare occurring on July 23, 2002. Gray bands show the search windows as same in Figure~\ref{fig:light_curves}. The vertical dotted lines show the peak timing for each channel.}\label{fig:light_10}
\end{figure}

\begin{figure}
\centerline{\includegraphics[width=0.8\textwidth,clip=]{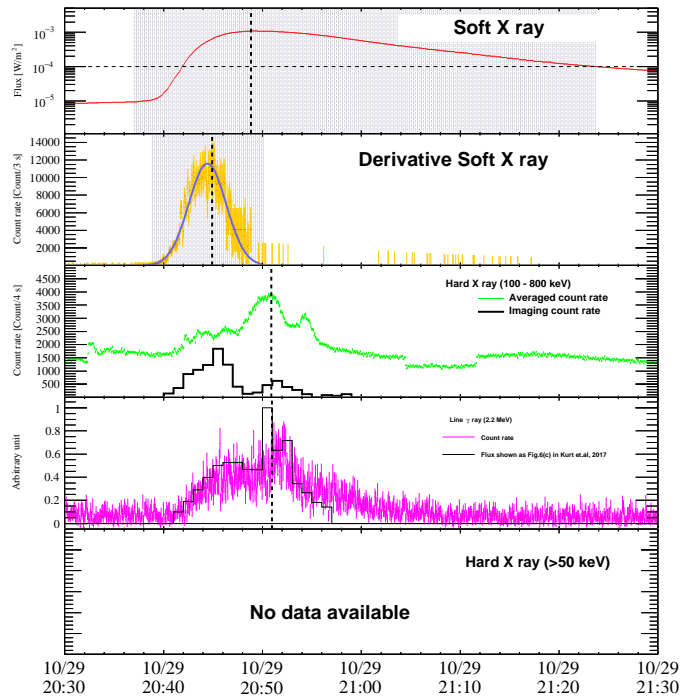}}
\caption{Light curves of the solar flare occurring on October 29, 2003. Gray bands show the search windows as same in Figure~\ref{fig:light_curves}. The vertical dotted lines show the peak timing for each channel. As described in the appendix text, both hard X-ray and line $\gamma$-ray are not used to determine the time window since their light curves were contaminated. The solid black line on third panel from top shows hard X-ray light curve using imaging method. The solid black line on fourth panel from top shows time profile of line $\gamma$-ray flux which is same as Fig.6(c) in ~\citealp{kurt2017}.}\label{fig:light_11}
\end{figure}

\begin{figure}
\centerline{\includegraphics[width=0.8\textwidth,clip=]{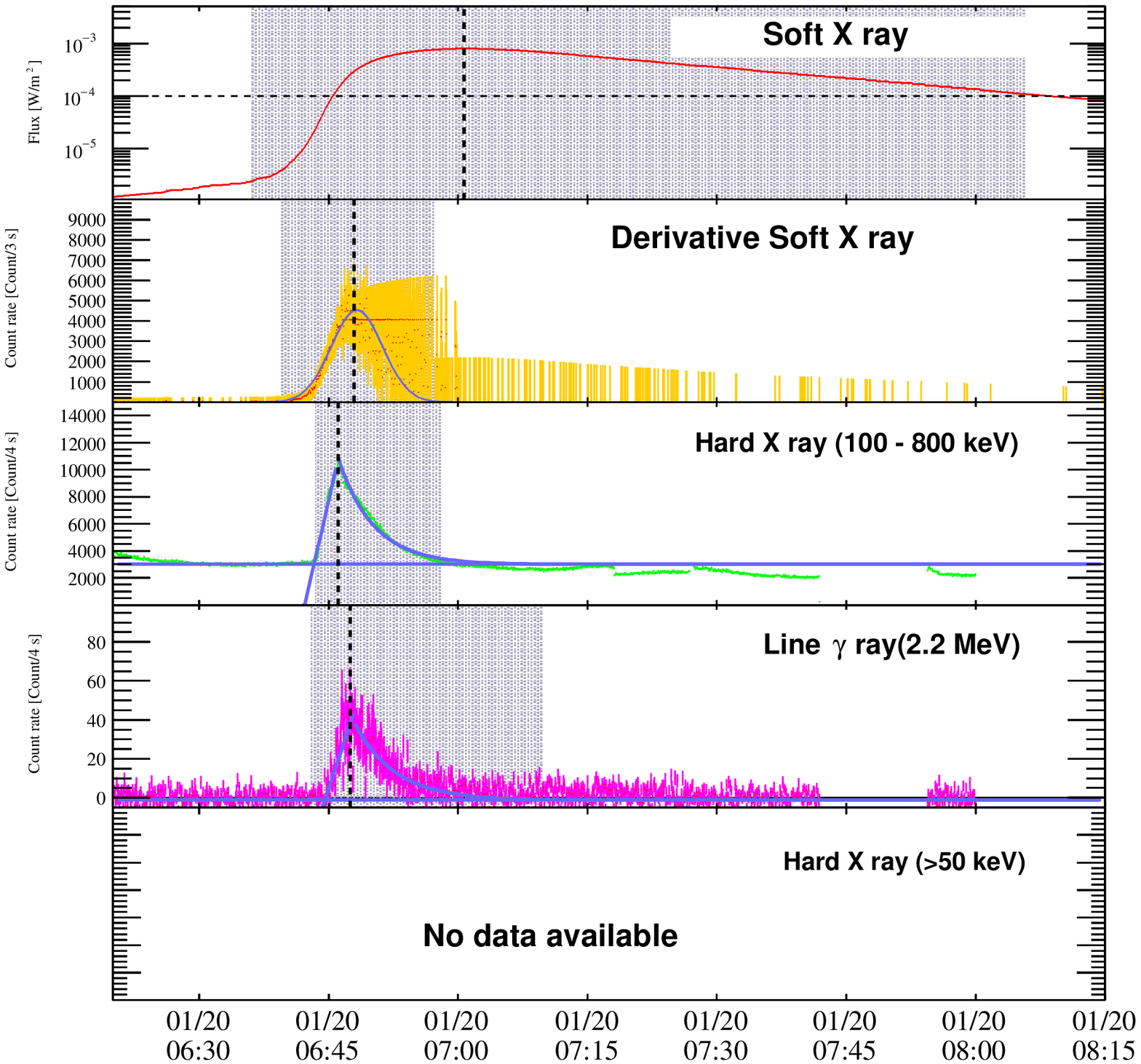}}
\caption{Light curves of the solar flare occurred on January 20, 2005. Gray bands show the search windows as same in Figure~\ref{fig:light_curves}. The vertical dotted lines show the peak timing for each channel.}\label{fig:light_12}
\end{figure}

As mentioned in section~\ref{sec:over}, the observation of $\gamma$-rays indicates the production of neutrinos via hadronic interactions, thus constituting the most important channel in this analysis. The present study captures four solar flares that include line $\gamma$-rays. In the main text, light curves for the solar flare occurring on November 2, 2003 are shown, because all channels are available for this event. Light curves of the other three flares are shown as follows: July 23, 2002 (Figure~\ref{fig:light_10}), October 29, 2003 (Figure~\ref{fig:light_11}), and January 20, 2005 (Figure~\ref{fig:light_12}).

\section{Comment on the Solar Flare Occurring on October 29, 2003}

The time profile of the flare occurring on October 29, 2003 is not similar to those of the other flares, because signal is contaminated with non-solar hard X-rays from magnetospheric origin. For only this flare, the peak timings of both hard X-rays and line $\gamma$-rays are delayed relative to that of soft X-ray as shown in Figure~\ref{fig:light_11}. The light curve of hard X-rays shows several minor peaks. To confirm the contaimination was caused by non-solar flare origin signal, we used spacial information from the imaging method to see the flare region on the surface of the Sun. The light curve with imaging method is shown in the third panel of Figure~\ref{fig:light_11}. The line $\gamma$--ray light curves shown in \citealp{kurt2017} is overlaid in the fourth panel of Figure~\ref{fig:light_11}.

\section{Light Curves for the Largest Solar Flare; November 4, 2003}

\begin{figure}
\centerline{\includegraphics[width=0.8\textwidth,clip=]{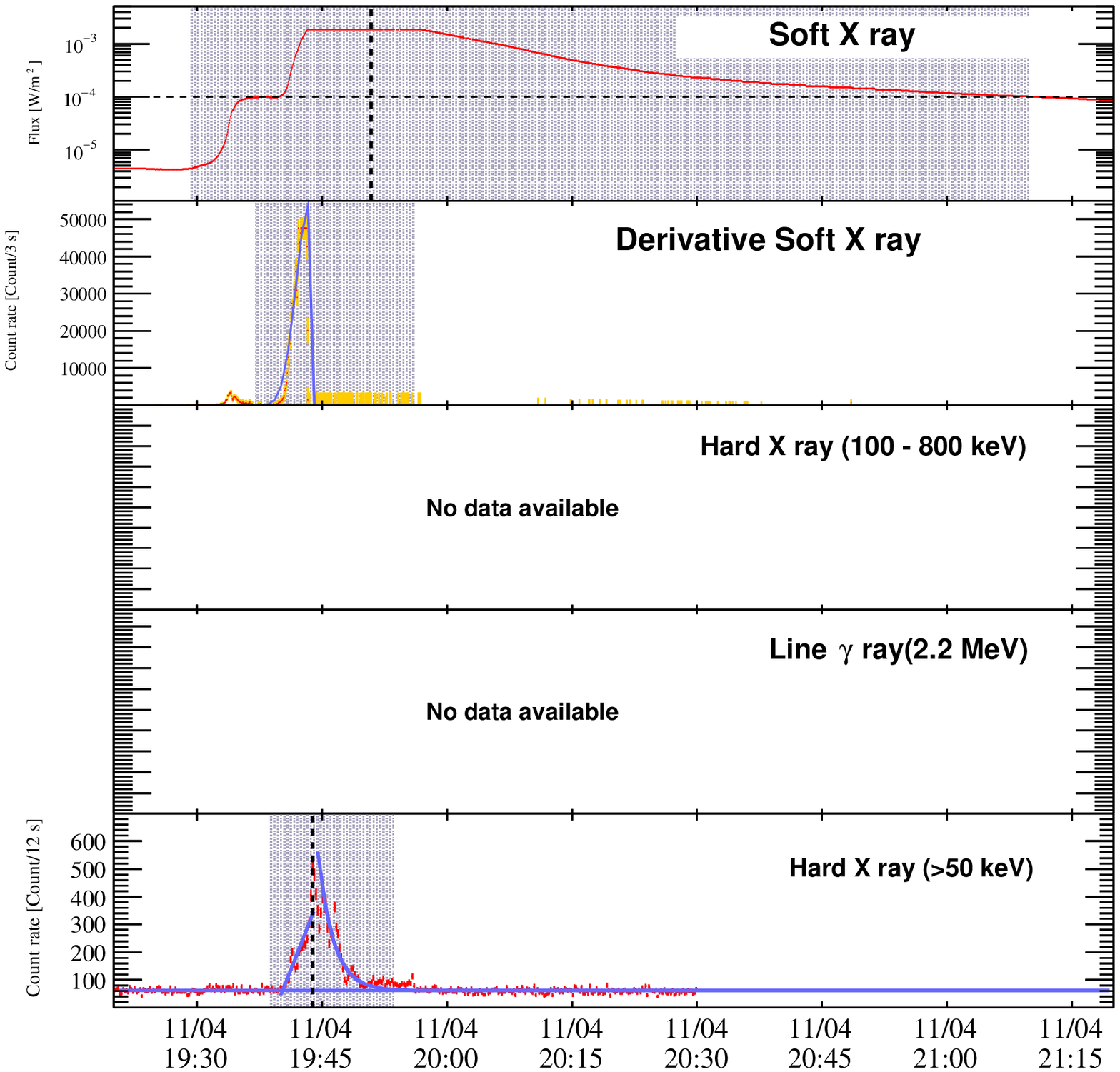}}
\caption{Light curves of the solar flare occurring on November 4, 2003. Gray bands show the search windows as same in Figure~\ref{fig:light_curves}. The vertical dotted lines show the peak timing for each channel. Because of high intensity, the soft X-ray flux was saturated.}\label{fig:light_x28}
\end{figure}

The largest solar flare on record occurred on November 4, 2003, with class X28.0. This flare attracts significant attention because of the presumably chance of solar flare neutrino detection. Figure~\ref{fig:light_x28} shows the light curves for this solar flare.

The GOES satellite instrument was saturated due to the high intensity of soft X-rays, and did not continue to record data for more than 15 min around 19:45--20:00. Because of this situation, we set the peak timing of soft X-ray for this flare at the middle point of the saturation phase, instead of the way outlined in the body of this paper. On the other hand, we could not set the peak timing of the derivative of soft X-rays due to the saturation. For this reason, we excluded this flare from comparison of peak timing with soft X-rays shown in  Figure~\ref{fig:diff_peak}. Moreover, we made a special treatment to set the end time of the derivative of soft X-rays at the end time of the saturation of soft X-rays.

Unfortunately, the RHESSI satellite did not record data related to this solar flare because it entered the Earth's shadow soon after the flare occurred.

%%%%%%%%%%%%%%%%%%%%%%%%%%%%%%%%%%%%%%%%%%%%%%%%%%%%%%%%%%%%%%
% Bibliography
%
% Using BibTeX
%
% \bibliographystyle{spr-mp-sola}
% \bibliography{<bib file>}
%
% Without BibTeX
% \begin{thebibliography}{}
% \bibitem[\protect\citeauthoryear{Author}{Year}]{key}
%   <bibliographical entry>
%
% \bibitem[\protect\citeauthoryear{}{}]{}
%
%
% \end{thebibliography}

\bibliographystyle{spr-mp-sola}
\bibliography{flare_draft_revis}

\end{article}
\end{document}